\documentclass[%
reprint,
superscriptaddress,
nofootinbib,
amsmath,amssymb,
aps,prx,
twocolumn,
floatfix,
]{revtex4-2}
\usepackage{tikz-cd}
\usepackage{braket}
\usepackage{relsize}
\usepackage{mathtools}
\usepackage{graphicx}
\usepackage{dcolumn}
\usepackage{bm}
\usepackage{hyperref}
\usepackage{cleveref}
\crefname{equation}{Eq.}{Eqs.}
\usepackage{comment}
\usepackage{float}
\usepackage{xcolor}
\usepackage{makecell}
\usepackage{enumitem}

\newcommand{\IQ}{Institut Quantique and Département de Physique,
  Université de Sherbrooke, Sherbrooke, Québec, J1K 2R1, Canada}
\newcommand{\Math}{Département de Mathématiques,
  Université de Sherbrooke, Sherbrooke, Québec, J1K 2R1, Canada}

\newcommand{\ha}{\hat{a}}
\newcommand{\had}{\hat{a}^\dag}
\newcommand{\balpha}{\boldsymbol{\alpha}}
\newcommand{\Pcode}{I}
\newcommand{\Icode}{I}
\newcommand{\Xcode}{X}
\newcommand{\Ycode}{Y}
\newcommand{\Zcode}{Z}

\newtheorem{proposition}{Proposition}
\begin{document}

\preprint{APS/123-QED}

\title{Error-Correction Properties of Bosonic Covariant Encodings}

\author{Frédéric St-Amand}
\email{frederic.st-amand@usherbrooke.ca}
\affiliation{\IQ}

\author{Jean-Philippe Burelle}
\affiliation{\Math}

\author{Baptiste Royer}
\affiliation{\IQ}

\date{\today}

\begin{abstract}
Bosonic codes offer a promising approach towards hardware-efficient fault tolerance, with many leading examples such as the cat-code and the GKP code organized by an underlying symmetry. We build on a representation-theoretic framework of quantum error correction to formalize the construction and analysis of multimode symmetric bosonic codes, focusing on finite groups. Here, a physical, logical representation and an initial seed state together define the codewords via a covariant encoding. A central observation is that QEC matrices can be viewed as morphisms of group representations, so that Schur's lemma forces entire irreducible representations blocks to vanish. We recover several known single- and two-mode bosonic codes and build and analyze the error protection capabilities of three codes based on finite subgroups of $SU(2)$. We analyze the impact of reducibility of the logical representation and optimize the seed state using the near-optimal fidelity.
\end{abstract}

\maketitle
\section{\label{sec:level1}Introduction}
Quantum error correction codes encode logical information in a subspace of a larger physical Hilbert space, protecting against specific types of errors. Such codes are typically designed around a particular physical error model but there is growing interest in constructions where the logical operations are easy to implement and the stabilizers are easy to measure. These properties are naturally supplied by symmetries. Separately, codes often require multiple physical qubits to encode a single logical qubit, leading to significant hardware overhead \cite{Corcoles2020, Fowler2012}. Bosonic codes address the hardware overhead challenge by encoding a logical qubit within the infinite-dimensional Hilbert space of a bosonic mode \cite{Cai2021}. This direct correspondence between physical object and logical qubit reduces the hardware overhead of quantum codes compared to qubit-register codes. In recent years, bosonic codes have emerged as a very active area of research, leading to the development of state-of-the-art quantum error-correcting codes. Notable examples include Gottesman-Kitaev-Preskill (GKP) codes \cite{Gottesman2001}, cat codes \cite{Leghtas2013}, rotation-symmetric bosonic codes \cite{Grimsmo2020} and quantum spherical codes \cite{Jain2024}.  

A common feature of error correction codes is the underlying symmetries, which can be systematically described using group and representation theory. Associating logical operations with symmetry transformations on the physical Hilbert space enables one to construct encodings that ensure a physical operation implements a logical operation \cite{Gross2021,Denys2023,Covariant2024,Leverrier2026,Ahmed2026}.
Denys and Leverrier~\cite{Covariant2024} leveraged symmetries of bosonic codes and studied the notion of a covariant encoding, in which a chosen physical representation of a finite group implements a target set of logical gates through simple physical operations~\cite{Denys2023, Covariant2024}. This perspective unifies known construction like the GKP code and the cat qudit encoding from an appropriate choice of group. It also provides a systematic approach to building new code including multimode codes built from orbits of coherent states. The representation-theoretic viewpoint was studied by Kubischta and Teixeira, who introduced the concept of intrinsic quantum codes~\cite{Kubischta2025}. 
There, the notion of code depth is introduced via representation theory, and error protection is directly related to the decomposition of errors into irreducible representations. 
Beyond theoretical appeal, bosonic codes are among promising experimental platforms for hardware-efficient error correction, with several experiments demonstrating quantum memory beyond the break-even point ~\cite{Ofek2016,Sivak2023, Ni2023,Brock2025}.

In this work, we leverage this symmetry-based approach with tools from representation theory to develop a general construction of covariant bosonic codes and analyze their error-correction properties. We apply this method to bosonic codes generated by finite subgroups of $SU(2)$ for single- and two-mode bosonic codes. Starting from a state in the physical Hilbert space, together with a physical and logical representation, we use the Jordan-Schwinger map to realize the group action on Fock space and the projection formulas of representation theory to construct explicit codewords. We then show that the transformation of photon loss errors under the group action, captured by decomposition of each error sectors into irreducible representations, controls which errors a code can detect and correct. Using this framework we analyze a family of single and two-mode codes, including the four-legged cat codes, the cyclic codes and new two-mode codes built from the quaternion group $Q_8$ and the binary octahedral group $2O$. For each code we quantify the performance against the amplitude damping channel using the quantum error correction (QEC) matrix. Restricting to coherent states, the initial seed state is optimized using the near-optimal fidelity.

This paper is organized as follows: \cref{sec:Prelim} establishes the necessary background, \cref{sec:generalConstruction} builds the general framework establishing our central observation that the QEC matrix is a morphism of group representations. We then apply this framework to single-mode codes in \cref{sect:singlemodecode} and to two-mode codes based on the quaternion group $Q_8$ and binary octahedral group $2O$ in \cref{sect:twomodecode}.
\section{Preliminaries}\label{sec:Prelim}
\subsection{Representation Theory}
The following results are standard in representation and group theory. For further details, we refer the reader to Ref.~\cite{Fulton2004}.

A \textit{representation} $\rho$ of a finite group $G$ on a finite-dimensional complex vector space $V$ is a homomorphism $\rho:G\longrightarrow GL(V)$. It is usual, when there's no ambiguity about the representation, to call $V$ the representation itself. If $V$ is also equipped with an inner product, a representation is said to be unitary if $\rho(g)$ is a unitary operator on $V$ for all elements of the group.
The \textit{kernel} of a representation is a normal subgroup of $G$ defined by 
\begin{align}
    \mathrm{ker}(\rho) = \{g\in G \mid \rho(g)=I\}.
\end{align}

A \textit{subrepresentation} of a representation $V$ is a vector subspace $W$ of $V$ which is invariant under $G$. If the only $G$-invariant subspace of $V$ are $\{0\}$ and $V$, then $V$ is said to be \textit{irreducible}. Any reducible representation $V$ of a compact group $G$ can be decomposed into irreducible components
\begin{align}
    V \cong \bigoplus_{i=1}^k V_i^{\oplus m_i},
\end{align}
where $V_i$ are irreducible representation of $G$ with multiplicity $m_i$. The subspace $V_i^{\oplus m_i}$ is referred to as the $i$th isotypic component of $V$.

Let $(\rho,V)$ and $(\sigma,W)$ be two representations of $G$. A \textit{morphism} $\varphi$ between representations is a linear map $\varphi : V\rightarrow W$ such that the following diagram commutes.
\[
\begin{tikzcd}
V \arrow[r, "\varphi"] \arrow[d,"\rho(g)"'] 
& W \arrow[d, "\sigma(g)"] \\
V \arrow[r, "\varphi"'] 
& W
\end{tikzcd}
\]

\textit{Schur's lemma}: Let $(\rho, V)$ and $(\sigma, W)$ be two irreducible representations of $G$ and $\varphi: V\rightarrow W$ a morphism of representations.
\begin{enumerate}
    \item Either $\varphi$ is an isomorphism or $\varphi = 0$.
    \item If $V = W$ then $\varphi = \lambda I$, for $\lambda\in\mathbb{C}$. 
\end{enumerate}

The \textit{dual} $\rho^*$ of a representation is defined as 
\begin{align}
    \rho^*:G&\longrightarrow GL(V^*)\\
    g&\longmapsto \rho(g^{-1})^t \nonumber
\end{align}
where $V^* = \mathrm{Hom}(V,\mathbb{C})$ is the dual vector space of linear forms on $V$.

The \textit{character} $\chi_V$ of a representation $V$ of $G$ is the complex-valued function on $G$ defined by
\begin{align}
    \chi_V(g) = \mathrm{Tr}(\rho(g)|_V).
\end{align}
Any representation of $G$ is completely determined up to isomorphism by its character.

The \textit{inner product} between the characters of two representation $\rho,\sigma$ of a finite group $G$ is defined as 
\begin{equation}
    \langle\chi_\rho,\chi_\sigma \rangle = \frac{1}{|G|}\sum_{g\in G}\overline{\chi_\rho}(g)\chi_\sigma (g).
\end{equation}
The characters form an orthonormal basis for the space of linear combination of characters.

The \textit{projection operator} onto an irreducible representation is given by
\begin{align}\label{Projection}
    P_{\sigma} = \frac{\mathrm{dim}(\sigma)}{|G|}\sum_G \overline{\chi_\sigma}(g)\rho(g). 
\end{align}

Let $G$ be a Lie group and $\mathfrak{g}$ the associated Lie algebra. A Lie algebra representation $\tau$ of $\mathfrak{g}$ on $V$ is a Lie algebra morphism $\tau : \mathfrak{g}\longrightarrow \mathfrak{gl}(V)$.

\subsection{Covariant Encoding}
Covariant encodings have been discussed in the contexts of QEC and metrology~\cite{Hayden2017,Faist2020,Liu2021,Kong2021,Zhou2021,Covariant2024,Kubischta2025}, and in this work we focus on encodings of finite-dimensional logical systems, expanding on Ref.~\cite{Covariant2024} which we summarize below.
Given a physical Hilbert space $\mathcal{H}_P$ and a logical Hilbert space $\mathcal{H}_L$, the encoding of a $d-$dimensional qudit within a physical system is given by an isometry $E:\mathcal{H}_L\longrightarrow\mathcal{H}_P$. Let $\rho$ and $\lambda$ be unitary representations of a given group $G$ in $\mathcal{H}_P$ and $\mathcal{H}_L$ respectively. We consider an encoding that is a morphism between the logical and physical representation of $G$, meaning that a physical operation implements a logical operation for all group elements, \emph{i.e.} $E\lambda(g) = \rho(g)E$ for all $ g \in G$. Let $V:\mathcal{H}_L\rightarrow \mathcal{H}_P$ be any linear map. In order to make $V$ a morphism of representations, one averages over the group elements 
\begin{align}\label{Eq:CovEnc}
    \tilde V \coloneqq \frac{1}{|G|}\sum_{G} \rho(g)V\lambda(g)^\dagger.
\end{align}
If $v = \mathrm{Tr}(V^\dag \tilde V)/d \neq 0$ then $V_G = v^{-1/2}\tilde V$ is a unitary representation of $G$ and we define $E = V_G$ to be the encoding. In what follows, we take $V=\ket{\Phi}\bra{\Psi}$ where $\ket{\Phi}\in\mathcal{H}_P$ and $\ket{\Psi}\in\mathcal{H}_L$. The codespace is then defined as $\mathcal{C}\coloneqq \mathrm{Im}(V_G)$.

\subsection{Quantifying Correctability}
The Knill-Laflamme condition gives a necessary and sufficient criterion for error correction of quantum codes given a set of errors $\{\hat L_k\}$. It is given by 
\begin{align}\label{Eq:KLC}
    \bra{i_L}\hat{L}_k^\dagger \hat{L}_l \ket{j_L} &= c_{kl} \delta_{ij},
\end{align}
where $\hat{L}_k$ are error operators and $\ket{i_L},\ket{j_L}$ is an orthonormal basis for the code space~\cite{Knill2000,Knill1997}. When the left hand side of \cref{Eq:KLC} is equal to zero, there exist a protocol to correct against such errors. More generally, it is useful to compute the QEC matrix $Q$ of a code for a given error channel defined as~\cite{Albert2018}
\begin{align}
    Q_{kl} = E^\dagger\hat{L}_k^\dagger \hat{L}_l E 
\end{align}
where $E$ is an isometry equivalent to projecting onto the code space and $Q_{kl}$ is a $d \times d$ matrix. The Knill-Laflamme condition can be restated as $Q_{kl} \propto E^\dagger E$ for all $k,l$.
The QEC matrix is a useful both as a numerical tool to quantify the degree to which it deviates from the Knill-Laflamme condition~\cite{Leung1997,Ng2010,Beny2010,Albert2018,Zheng2024,Li2025}, and as a visual tool to illustrate the effect of errors in the code space~\cite{Albert2018}. For logical qubits ($d=2$), the matrix $Q_{kl}$ can be decomposed into
\begin{align}
    Q_{kl} = c_{kl}\Pcode+x_{kl}\Xcode+y_{kl}\Ycode+z_{kl}\Zcode,
\end{align}
where coefficients are given by
\begin{align}\label{QECMCoef}
    [c,x,y,z]_{kl} = \frac{1}{2}\mathrm{Tr}\{[\Pcode,\Xcode,\Ycode,\Zcode]E^\dag \hat{L}_k^\dagger \hat{L}_l E\}.
\end{align}

\subsection{Fidelity of an Encoding}
An important metric of a quantum error correction code under a noise channel is the fidelity which, given a physical quantum channel or a composition of them, gives a measure of the distortion on the logical states. Let $\mathcal{W}$ be a quantum channel. The channel fidelity is defined as~\cite{Nielsen_Chuang_2010}
\begin{align}\label{Eq:ChannelFidelity}
    F(\mathcal{W}) \coloneqq \bra{\psi}\mathcal{W\otimes}\mathcal{I}(\ket{\psi}\bra{\psi})\ket{\psi}
\end{align}
where $\ket{\psi}$ is a purified maximally mixed state and $\mathcal{I}$ is the identity channel on a reference system. Our interest lies in quantifying the preservation of logical information, and consequently we focus on logical channels. Given a noise channel $\mathcal N$ on a physical Hilbert space $\mathcal H_P$, we obtain a logical channel
\begin{equation}
    \mathcal W = \mathcal R \circ \mathcal N \circ \mathcal E,
\end{equation}
where $\mathcal{E}$ is the encoding channel and $\mathcal R$ is a recovery channel. To quantify the robustness of the encoding $\mathcal E$ against the noise channel $\mathcal N$ without relying on a specific recovery, one option is to use the optimal fidelity
\begin{equation}
    F_{\mathrm{opt}}(\mathcal W) = \max_{\mathcal R}F(\mathcal W).
\end{equation}
Although this optimization can be done in principle, its computational cost scales as $\Tilde{\mathcal{O}}((d_L N)^{5.246})$ where $d_L$ is the dimension of the logical Hilbert space and $N$ is the dimension of the physical Hilbert space~\cite{Jiang2020}. To avoid this high numerical cost we instead use the near-optimal fidelity as a figure of merit for our code whose computational cost scales as $\mathcal{O}((d_L N_k)^3)$ where $N_k$ is the number of Kraus operators in the calculation~\cite{Zheng2024}. This metric can be computed directly from the QEC matrix, and is defined as
\begin{align}\label{NearOptFid}
    \Tilde{F} = \frac{1}{d_L^2}\left\Vert\mathrm{Tr}_L(\sqrt{Q})\right\Vert^2_F
\end{align}
where $||.||_F$ denotes the Frobenius norm, $Q$ is the QEC matrix and $\mathrm{Tr_L}(B)$ is the partial trace over the codespace indices.  

\subsection{Jordan-Schwinger Representation}
The \textit{Jordan-Schwinger} map gives a way of representing Lie algebra elements as a function of bosonic or fermionic creation and annihilation operators~\cite{Schwinger1952, Dubus2024}. We denote by $\ha_j,\had_j$ the bosonic annihilation and creation operators of mode $j$, respectively. 

Let $G$ be a Lie group, $\mathfrak{g}$ the associated Lie algebra and $\mathcal{F}$ the bosonic Fock space over $\mathbb{C}^N$ equipped with the creation/annihilation operators $\hat{a}_1^\dagger,...,\hat{a}_N^\dagger,\hat{a}_1,...,\hat{a}_N$ satisfying the canonical commutation relations $[\hat{a}_i,\hat{a}_j^\dagger]=\delta_{ij}, [\hat{a}_i,\hat{a}_j]=[\hat{a}_i^\dagger,\hat{a}_j^\dagger]=0$. The Jordan-Schwinger representation is defined as
\begin{align}
   \mathrm{js}: \mathfrak{gl}(N,\mathbb{C}) &\longrightarrow \mathfrak{gl}(\mathcal{F})\\ 
   M &\longmapsto \sum_{ij} \hat{a}_i^\dagger M_{ij}\hat{a}_j.
\end{align}
Accordingly, we can lift the Jordan-Schwinger Lie algebra representation to a group representation JS with action on a N-mode coherent state $\ket{\balpha}$ given by
\begin{equation}\label{eq:JSgroupRep}
    \mathrm{JS}(U)\ket{\balpha} = \ket{U\balpha},
\end{equation}
where $U \in U(N)$, JS$(U)\in U(L^2(\mathbb R^N))$ and the action of $U$ on $\balpha \in \mathbb C^N$ inside the ket on the right-hand-side is the standard matrix-vector product. Note that for any unitary operator $U$, $\mathrm{JS}(U)$ commutes with the total photon number operator $\hat n = \sum_{j=1}^n \hat a^\dag_j \hat a_j$.

\subsection{Oscillator errors}\label{subsect:OscErrors}
Since our code construction are exclusively bosonic codes, the error model we study is the amplitude damping channel, for example photon losses when the bosonic modes are realized by microwave cavities.

Let us define an error $\boldsymbol{k} = (k_1,k_2,...)$ of order $K=|\boldsymbol{k}|=\sum_i^N k_i$, with each $k_j \in \mathbb N$, as
\begin{align}
    \hat{a}^{\boldsymbol{k}} &= \prod_{i=1}^N \hat{a}_i^{k_i}.
\end{align}
In other words, the error $\hat a^{\boldsymbol{k}}$ has the effect of removing $k_j$ photons on mode $j$.

The Kraus operators associated to the amplitude damping channel are not directly given by photon loss operators, but rather the
operators in the Kraus operator-sum decomposition for a single mode are~\cite{Kraus1983,Chuang1997}
\begin{equation}
    \hat L_{k} = \left(\frac{\gamma}{1 -\gamma}\right)^{k/2} \frac{\ha^{k}}{\sqrt{k!}}(1 - \gamma)^{\hat n/2},
\end{equation}
for a photon loss rate $\kappa$ and a time $t$, the dimensionless parameter $\gamma$ corresponds to the probability for each photon to be lost, $\gamma = 1 - \exp(-\kappa t)$. Assuming that the parameter $\gamma$ is identical for all modes, we can label these operators in a manner similar to photon loss operators, with the relation
\begin{equation}\label{eq:EvsL}
    \hat L_{\boldsymbol{k}} = \left(\frac{\gamma}{1 -\gamma}\right)^{|\boldsymbol{k}|/2}\left(\prod_{j=1}^N \frac{1}{\sqrt{k_j!}}\right) \hat a^{\boldsymbol{k}}(1 - \gamma)^{\hat n/2},
\end{equation}
The effect of a Kraus operator on a single-mode coherent state is given by
\begin{equation}
    \label{eq:EffectOfKrausOpOnCoherentState}
    \hat L_{\boldsymbol{k}} \ket{\boldsymbol{\alpha}} = \gamma^{|\boldsymbol{k}|/2}e^{-\gamma|\boldsymbol{\alpha}|^2/2}\left(\prod_{j=1}^N \frac{\alpha_j^{k_j}}{\sqrt{k_j!}}\right)\ket{\sqrt{1-\gamma}\boldsymbol{\alpha}}.
\end{equation}

\section{General construction}\label{sec:generalConstruction}
In this section, we expand on previous work on covariant encoding~\cite{Covariant2024,Kubischta2025,Jain2024,Kubischta2025}, and discuss their general error-correction capabilities. We first present covariant encodings in a more general setting. Then we discuss the errors they correct, and comment about the choice of initial physical state that generates the encoding.

\subsection{Encoding}\label{subsec:encoding}
The data required for the code construction can be summarized as follows:
\begin{enumerate}[label=(\arabic*)]
    \item Choose an abstract group $G$. 
    \item Choose a unitary representation $\rho$ of the group $G$ to define the physical representation.
    \item Choose a $d$-dimensional logical representation $\lambda$ of $G$.
    \item Choose an initial state in the physical Hilbert space of interest with nonzero projection onto the isotypic components of $\lambda$ in $\rho$, which fixes a specific realization of $\lambda$ as a subrepresentation and thereby defines a covariant isometry.
\end{enumerate}
We comment below on each of the choices above.

(1) In this work, we focus on finite subgroups of $U(1)$ and $SU(2)$ to encode quantum information in one or two harmonic oscillator modes.

(2) To define the physical representation $\rho$ of $G$, we first choose a N-dimensional faithful unitary representation $\rho_f$ of $G$.
The choice $\rho_f$ fixes the representation of $G$ on the one-excitation subspace of $\mathcal H_P$, which for a two-mode system corresponds to span$\{\ket{1,0},\ket{0,1}\}$. This representation can then be extended to the full Hilbert space by taking a direct sum of symmetric products of $\rho_f$ for the bosonic systems. That process yields a representation $\rho$ of $G$ on the full Fock space that is equivalent to the Jordan-Schwinger (group) representation defined in \cref{eq:JSgroupRep}. From here on out it is understood that applying an element $\rho(g)$ on a state is the composition $(JS\circ\rho_f)(g)$.

We note that choosing a representation $\rho_f$ that is not faithful is equivalent to choosing an initial group $H$ in step (1) that is a quotient of $G$, meaning that the faithfulness condition on $\rho_f$ ensures that the physical representation is injective without reducing the generality of the code construction. 

The action of $\rho$ of $G$ on $\mathcal{H}_P$ defines a natural action via conjugation on the space of operators on $\mathcal{H}_P$ given by
\begin{align}\label{Transform}
    \sigma(g)\cdot \hat{O} = \rho(g)\hat{O}\rho^\dagger(g), \;\forall \, g\in G,
\end{align}
where $\hat{O}$ is an operator acting on $\mathcal{H}_P$. Individual errors are expressed as operators on $\mathcal H_P$, and the representation $\sigma$ yields a lot of structure that we explore in \cref{subsect:GenConErrorCorrection}.

(3) Having defined the action of $G$ on the physical Hilbert space $\mathcal H_P$, we now choose the action on the abstract logical Hilbert space $\mathcal H_L$ via the representation $\lambda$. In this work, we focus on encoding logical qubits, \emph{i.e.} two-dimensional logical representations $\lambda$.
For a given group $G$, several choices of logical representation are possible, which can be obtained by listing all the different (potentially reducible) two-dimensional representations of $G$.

The kernel of each logical representation, $N = \mathrm{ker}(\lambda)$, defines a normal subgroup $N\unlhd G$, which in turn determines the stabilizer subgroup of the code $S\coloneqq \mathrm{ker}(\mathbb{P}[\lambda])$ where $\mathbb P [.]$ denotes the projective representation. We then have the following relation between the three groups $N \unlhd S \unlhd G$. The logical group is given by the quotient group $G/N$. The set of resulting logical gates $L$ is obtained after removing the effect of global phases, \emph{i.e.} we get $L \cong \mathrm{Im} (\mathbb P[\lambda])\cong G/S$. The quotient group $S/N$ constitutes the group of global phases; a finite cyclic subgroup of $U(1)$. It must be pointed out that the so-called projective representation is not a true representation. In fact, the projective representation is a morphism defined as $\mathbb{P}[\lambda]:G\longrightarrow PGL(V)\cong GL(V)/Z$ where $Z$ is the center of $GL(V)$. Therefore, $\mathbb{P}[\lambda]$ does not result in individual matrices but equivalence classes of matrices differing by a global phase. As a result, it is worthwhile to make explicit the difference between $S$ and $N$ when using properties based on representations. Furthermore, in stabilizer code theory the codespace is usually defined as the +1 eigenspace of the stabilizer group. However, as defined above $S$ can contain elements that leave the codespace invariant up to a global phase, for example $\rho(s)\ket{\psi} = -\ket{\psi}$. Typically, that difference is unimportant as one can redefine the stabilizers such that the code space is in the +1 eigenspace of the stabilizers, for example $\rho(s) \rightarrow -\rho(s)$. However, in the current construction $-\rho(s)$ may not be in the image of $\rho$, such that we find it more convenient to extend the definition of the stabilizer group $S$.

Ideally, one chooses a logical representation that balances logical operations and stabilizer operations. The former allow easier logical control of the qubit, while the latter allow better and easier detection of errors. Note however that, as we show below,  it is possible to obtain a code with correctable errors even if the stabilizer group is trivial. Moreover, with the current construction, the codewords defined by $V_G\ket{0/1}$ are automatically orthogonal. Codewords associated with non-isomorphic irreducible representations are automatically orthogonal under any
$G$-invariant inner product (Schur's orthogonality relations) and within a single isotypic component, orthogonality is fixed by the choice of basis in $\mathcal H_L$.

(4) Once an abstract logical representation $\lambda$ has been chosen, we aim to define an isometry between the abstract logical space $\mathcal H_L$ and the physical Hilbert space $\mathcal H_P$. To do so, one chooses a subrepresentation of $\rho$ that is isomorphic to $\lambda$. To that end, it is convenient to choose an initial physical state $\ket{\Phi} \in \mathcal H_P$ and project onto the desired representations using \cref{Eq:CovEnc}. In this work we focus on initial $N$-mode coherent states, $\ket{\Phi} = \ket{\balpha}$, leading to multimode generalizations of cat codes~\cite{Dodonov1974,Leghtas2013}.

In the case where $\lambda$ is chosen to be a direct sum of one-dimensional representations, say $\lambda \cong \rho_m \oplus \rho_n$, where $m\neq n$, we find the following encoding formula equivalent but easier to work with than \cref{Eq:CovEnc},
\begin{equation}
\label{eq:VgVersionSumOf1Dreps}
    V_G = P_{\rho_m}\frac{\ket{\balpha}\bra{0}}{\sqrt{\bra{\balpha}P_{\rho_m} \ket{\balpha}}}  + P_{\rho_n}\frac{\ket{\balpha}\bra{1}}{\sqrt{\bra{\balpha}P_{\rho_n} \ket{\balpha}}}  .
\end{equation}
where $\ket{\balpha} \in \mathcal{H}_P$. \Cref{Eq:CovEnc} is recovered with the choice $\ket{\Psi} \propto \ket{0}/\sqrt{\bra{\balpha}P_{\rho_m} \ket{\balpha}} + \ket{1}/\sqrt{\bra{\balpha}P_{\rho_n} \ket{\balpha}}$. The above formula fixes the logical gates to be diagonal in the computational basis without loss of generality.

As shown below, possible choices of logical representations includes ones of the form $\lambda\cong\rho_n\oplus\rho_n$, in which case \cref{eq:VgVersionSumOf1Dreps} needs to be modified (see \cref{app:ModEncoding}).

\subsection{Error correction}\label{subsect:GenConErrorCorrection}
\subsubsection{Physical Errors}
Using representation theory as a tool to understand how operators transform under the action of a finite group provides significant analytical simplification when analyzing error-correction properties of the code.

Importantly, the physical representation $\rho$ preserves the number of excitation, and the associated action of $G$ via the representation $\sigma$ on $\hat a^{\boldsymbol{k}}$ yields a superposition of errors $\hat a^{\boldsymbol{k}'}$ with $|\boldsymbol{k}| = |\boldsymbol{k}'| = K$. One can therefore analyze the action of $\sigma$ on the vector spaces generated by operators of same order $K$.  We denote the set of all photon-loss errors of order $K$ as $\mathcal A_K \coloneqq \{\hat{a}^{\boldsymbol{k}}\}_{|\boldsymbol{k}|=K}$, and the restriction of $\sigma$ to $\mathrm{Span}\{\mathcal A_K\}$ as $\sigma|_K$. We also define the set of all photon-gain operators $(\hat a^{\boldsymbol{k}})^\dag$ of order $|\boldsymbol{k}| = K$ as $\mathcal A^\dag_K$ and the restriction of $\sigma$ to the vector space $\mathrm{Span}\{\mathcal A_K^\dag\}$ as $\sigma|_{K^\dag}$.
\begin{proposition}
For operators of order 1, \emph{i.e.} where a single photon is lost, the action of $\sigma$ is isomorphic to the dual of the representation $\rho_f$ on the single-excitation subspace, 
\begin{equation}
    \begin{aligned}
    \sigma\big|_{K=1}&\cong \rho_f^*, \\ 
    \sigma\big|_{K^\dag=1}&\cong \rho_f, \\ 
    \end{aligned}
\end{equation}
Written differently, for a vector element $\boldsymbol{a} \in \mathrm{Span}\{\mathcal A_1\}$ written $\boldsymbol{a} = \sum x_k \hat a_k$, we have 
\begin{equation}
    \sigma\big|_{K=1}(g) \boldsymbol{a}  = \rho(g) \boldsymbol{a} \rho^\dagger(g)= \sum_{j,k}[\rho_f^*(g)]_{jk} x_k \hat a_j.
\end{equation}
\end{proposition}
We refer to \cref{app:Proposition1} for the proof of that statement. Single photon annihilation operators acting on superposition of single photon states can be viewed as linear functionals, i.e., linear maps from a vector space to the underlying field that preserve vector addition and scalar multiplication. Since bosonic annihilation operators commute between modes, we obtain
\begin{equation}
    \begin{aligned}
    \sigma\big|_{K} &\cong \mathrm{Sym}^K(\rho_f^*), \\
    \sigma\big|_{L^\dag}  &\cong \mathrm{Sym}^L(\rho_f^*)^* \cong \mathrm{Sym}^L(\rho_f).
    \end{aligned}
\end{equation}
A similar argument can be made for monomials $(\hat{a}^{\boldsymbol{l}})^\dagger \hat{a}^{\boldsymbol{k}}$, with the action of $G$ yielding a superposition of terms all with $L = |\boldsymbol{l}|$ and $K = |\boldsymbol{k}|$ constant. As a result, the restriction $\sigma|_{L^\dag,K}$ of $\sigma$ to the span of such monomials, $\mathrm{Span}\{\mathcal{A}_L^\dagger \mathcal{A}_K\}$, is isomorphic to 
\begin{equation}
    \sigma\big|_{L^\dag,K} \cong \mathrm{Sym}^L(\rho_f)\otimes \mathrm{Sym}^K(\rho_f^*).
\end{equation}

One can decompose specific error sectors into irreducible representation components,
\begin{equation}
    \sigma|_{L^\dag,K} \cong \bigoplus_j \rho_j^{m_j(L,K)},
\end{equation}
where $j$ indexes the representations of $G$ and $m_j$ is the multiplicity of that representation, a number which depends on $L$ and $K$. This decomposition procedure is computationally straightforward and provides substantial insight into the error-correction properties of the code. Specific examples of such calculations are made explicit in \cref{sect:singlemodecode,sect:twomodecode}.  

Since all operators $\rho(g)$ commute with the total photon number $\hat n$, the action of $\sigma(g)$ on the set of photon loss Kraus operators leaves the order of the error $|\boldsymbol{k}|$ invariant. Defining the set of all order-$K$ Kraus operators $\mathcal{L}_K\coloneqq \{\hat{L}_{\boldsymbol{k}}\}_{|\boldsymbol{k}|=K}$, we obtain a non-unitary representation $\tilde \sigma|_{K}$ of $G$ restricted to the vector space $\mathrm{span}\{\mathcal{L}_K\}$ given by
\begin{equation}
\begin{aligned}
    \rho(g) \boldsymbol{e}\rho^\dag(g) &= P \sigma(g) P^{-1}\cdot \boldsymbol{e},\\
    &= \tilde \sigma(g) \cdot\boldsymbol{e},
\end{aligned}
\end{equation}
where $\boldsymbol{e} \in \mathrm{span}\{\mathcal{L}_K\}$ and $P$ is a similarity transformation defined by
\begin{equation}
    P \cdot \hat L_{\boldsymbol{k}} = \left(\Pi_{j=1}^N \frac{1}{\sqrt{k_j!}}\right) \hat L_{\boldsymbol{k}}.
\end{equation}
Since $1!=0! = 1$, in the $K=1$ space we have $P=I_N$ and correspondingly $\tilde \sigma|_1 = \sigma|_1 = \rho_f^*$. Through a similar calculation, we compute that the action of $\sigma$ restricted to $\mathrm{span}\{\mathcal L^\dag_K\}$ is isomorphic to $\sigma|_{K^\dag}$ and that
\begin{equation}
    \tilde \sigma\big|_{L^\dag,K} \cong \sigma\big|_{L^\dag,K} \cong \mathrm{Sym}^L(\rho_f)\otimes \mathrm{Sym}^K(\rho_f^*).
\end{equation}

As the calculation above shows, the space of Kraus operators for the photon loss channel, $\mathrm{span}\{\mathcal L_{L}^\dag \mathcal L_K\}$, transforms as a non-unitary representation isomorphic to the simpler space of photon loss operators, $\mathrm{span}\{\mathcal{A}_L^\dagger \mathcal{A}_K\}$. For all analysis involving the presence or absence of some irreducible representation in the spaces of errors, one can therefore carry the calculation with the simpler photon loss operators $\{\hat a^{\boldsymbol{k}}\}$, and extend the conclusions to the Kraus operators of the photon loss channel.

The QEC matrix associated to the photon loss channel can be related with that of the photon loss operators even more precisely, meaning that from the QEC matrix computed from the photon loss operators, the QEC matrix of the photon loss channel Kraus operators can be computed. 
We focus on encoding logical qubits and we distinguish the cases where (A) the logical representation is a sum of one-dimensional irreducible representations of $G$, say $\lambda \cong \rho_m \oplus \rho_n $ and (B) the logical representation is isomorphic to a two-dimensional irreducible representation of $G$. 

In case (A), we combine \cref{eq:EffectOfKrausOpOnCoherentState,eq:VgVersionSumOf1Dreps} to obtain
\begin{equation}
    \hat L_{\boldsymbol{k}}V_G(\balpha) = \eta_{\boldsymbol{k}} \hat a^{\boldsymbol{k}}V_G(\balpha')W(\balpha)
\end{equation}
with a scalar factor
\begin{equation}
    \eta_{\boldsymbol{k}} = \left(\frac{\gamma}{1 -\gamma}\right)^{|\boldsymbol{k}|/2} e^{-\gamma |\boldsymbol{\alpha}|^2} \left(\Pi_{j=1}^N \frac{1}{\sqrt{k_j!}}\right)
\end{equation}
and a correction
\begin{equation}
\begin{aligned}
    W(\balpha) ={}& \frac{I}{2}\left(\sqrt{\frac{\bra{\balpha'}P_{\rho_m} \ket{\balpha'}}{\bra{\balpha}P_{\rho_m} \ket{\balpha}}} + \sqrt{\frac{\bra{\balpha'}P_{\rho_n} \ket{\balpha'}}{\bra{\balpha}P_{\rho_n} \ket{\balpha}}}\right) \\
    &+ \frac{Z}{2}\left(\sqrt{\frac{\bra{\balpha'}P_{\rho_m} \ket{\balpha'}}{\bra{\balpha}P_{\rho_m} \ket{\balpha}}} - \sqrt{\frac{\bra{\balpha'}P_{\rho_n} \ket{\balpha'}}{\bra{\balpha}P_{\rho_n} \ket{\balpha}}}\right),
\end{aligned}
\end{equation}
where we have defined $\balpha' := \sqrt{1-\gamma}\balpha$ for conciseness.

This relation allows to compute the QEC matrix for the photon loss channel and an initial state $\balpha$ using the QEC matrix computed from the simpler photon loss operators $\{\hat a\}$ on a smaller code $\balpha' = \sqrt{1-\gamma}\balpha$,
\begin{equation}
\begin{aligned}
    V_G^\dag(\balpha) \hat L^\dag_{\boldsymbol{l}}& \hat L_{\boldsymbol{k}} V_G(\balpha) =\\
   & W^\dag_{\mathcal{C}}(\balpha)V_G^\dag(\balpha') (\hat a^{\boldsymbol{l}})^\dagger \hat a^{\boldsymbol{k}} V_G(\balpha')W_{\mathcal{C}}(\balpha) \eta_{\boldsymbol{l}} \eta_{\boldsymbol{k}}.
    \end{aligned}
\end{equation}
In particular, for the ``no error" block $Q_{0,0}$ of the QEC matrix, we obtain
\begin{equation}
    Q_{0,0} = W^\dagger_{\mathcal{C}}(\balpha)W_{\mathcal{C}}(\balpha) e^{-2\gamma |\alpha|^2},
\end{equation}
making explicit the fact that the ``no-jump" part of the channel causes logical dephasing errors. 

In case (B), where the logical representation is isomorphic to a two-dimensional irreducible representation, we get
\begin{equation}
    \hat L_{\boldsymbol{k}}V_G(\balpha) = \eta_{\boldsymbol{k}} \hat a^{\boldsymbol{k}}V_G(\balpha') \frac{\mathrm{Tr}\left[\ket{0}\bra{\balpha'}\tilde V(\balpha')\right]}{\mathrm{Tr}\left[\ket{0}\bra{\balpha}\tilde V(\balpha)\right]},
\end{equation}
which again allow to compute the elements of the QEC matrix using the photon loss operators. Notably, this equation shows that $\hat L_{\boldsymbol{k}}V_G(\balpha) \propto \hat a^{\boldsymbol{k}}V_G(\balpha')$, such that a code that can perfectly correct a set of photon loss operators can also correct the corresponding operators of the full channel. In particular, the ``no error" case is exactly correctable in these cases, \textit{i.e.} the $Q_{0,0}$ block of the QEC matrix is proportional to a logical identity.

\subsubsection{Representation theory of the QEC matrix}
In this section we present the QEC matrix in the representation theoretic framework. Every sesquilinear form is represented by a unique operator $M$ through
\begin{align}
    Q_M(\ket{\alpha},\ket{\beta}) = \bra{\alpha} M \ket{\beta}.
\end{align}
The assignment $M\mapsto Q_M$ is a linear isomorphism such that the two can be identified together. For a qubit, $Q_M$ is a 2 by 2 matrix. 
The space of sesquilinear forms over $\mathcal{C}$ is naturally a vector space of dimension $\mathrm{dim}(\mathcal{C})^2$ isomorphic to $\mathrm{Sesq}(\mathcal{C})\cong \overline{\mathcal{C}}^*\otimes\mathcal{C}^*$, where the inner product supplies the antilinear isomorphism $\overline{\mathcal{C}}^*\cong \mathcal{C}$ and $\mathcal{C^*}\cong \overline{\mathcal{C}}$. As mentioned above, we often denote the vector space as the representation, and we use interchangeably $\overline{\mathcal{C}}^*\otimes\mathcal{C}^*$ and $\overline{\lambda}^*\otimes\lambda^*$.

From a pair of photon loss Kraus operators $\hat{L}_{\boldsymbol{k}}, \hat{L}_{\boldsymbol{l}} \in \mathrm{Sym}(\mathcal{H}_P^*)$, a sesquilinear form on $\mathcal{H}_L$, isomorphic to a sesquilinear form on $\mathcal{C}$, is defined as 
\begin{align}
    Q_{\boldsymbol{k},\boldsymbol{l}}(\ket{\mu},\ket{\nu}) \coloneqq  \bra{\mu}V_G^\dagger\hat{L}_{\boldsymbol{k}}^\dagger \hat{L}_{\boldsymbol{l}} V_G\ket{\nu},
\end{align}
where the indices $\boldsymbol{k},\boldsymbol{l}$ specify the matrix $M$ and $\ket{\mu},\ket{\nu}\in \mathcal{H}_L$. This in turn defines an action $\star$ of $G$ on $\mathrm{Sesq}(\mathcal{C})$ given by 
\begin{equation}
g\star Q_{\boldsymbol{k},\boldsymbol{l}}(\ket{\mu},\ket{\nu}) = Q_{\boldsymbol{k},\boldsymbol{l}}(\lambda(g)^\dagger\ket{\mu},\lambda(g)^\dagger\ket{\nu}).
\end{equation}
From the identification above, the QEC matrix is a morphism of group representations. This statement is made explicit in the following proposition.
\begin{proposition}\label{prop:QECmatrixIsMorphism} Given an error model and a subspace of operators $L$ where $L$ also needs to be a subrepresentation of $G$ over the physical error space, the map 
\[ 
    \Lambda : \overline{L}^*\otimes L^* \longrightarrow \mathrm{Sesq}(\mathcal{C})
\]
defined by $\Lambda(\hat{L}_{\boldsymbol{k}}\otimes\hat{L}_{\boldsymbol{l}})\coloneqq Q_{\boldsymbol{k},\boldsymbol{l}}$ is a morphism of representations. The induced action ($\star$) on the QEC matrix is conjugation by the logical representation
\begin{equation}
    g \star Q_{\boldsymbol{k},\boldsymbol{l}} = \lambda(g) Q_{\boldsymbol{k},\boldsymbol{l}}\lambda(g)^\dag.
\end{equation}
\end{proposition}
We refer to \cref{app:ProofPropo3} for the proof. See also Ref.~\cite{Kubischta2025} for a similar notion with intrinsic quantum codes. 

For any representation of $G$ on the code space $\mathcal C$, a decomposition of $\overline{\mathcal{C}}^*\otimes\mathcal{C}^*$ into irreducible representations always involves at least one trivial representation which corresponds to the logical identity matrix $\Icode$ as $I$ is left invariant by conjugation of all group elements.

Using \cref{prop:QECmatrixIsMorphism} in conjunction with Schur's lemma, there are few cases where the error-correction analysis can be made simpler.
\begin{enumerate}[label=(\roman*)]
    \item If there are no irreducible factors of the representation $\tilde \sigma_{L^\dag,K}$ of $G$ that appear in $\overline{\mathcal{C}}^*\otimes\mathcal{C}^*$ : we can conclude without further calculations that $Q_{\boldsymbol{l},\boldsymbol{k}} = 0$ for all $|\boldsymbol{l}|=L,|\boldsymbol{k}|=K$.
    \item If the only irreducible factor of $\tilde \sigma_{L^\dag,K}$ that appears in $\overline{\mathcal{C}}^*\otimes\mathcal{C}^*$ is the trivial representation \emph{and} $\overline{\mathcal{C}}^*\otimes\mathcal{C}^*$ carries a single trivial representation: The errors are necessarily correctable.
\end{enumerate}
Note that when $\lambda$ is reducible, case (ii) does not happen, as $\overline{\mathcal{C}}^*\otimes\mathcal{C}^*$ contains at least two copies of the trivial representation as shown in \cref{prop:lambdaRep} below. 

The following proposition shows how to compute the multiplicity of the trivial irreducible representation in the decomposition of $\mathrm{Sesq}(\mathcal{C})$.

\begin{proposition}\label{prop:lambdaRep}
The multiplicity of the trivial representation in the decomposition of $\mathrm{Sesq}(\mathcal{C})$ into irreducible factors is given by $\sum_i m_i^2$ where $i$ indexes the irreducible representations in $\mathcal C$ and $m_i$ are the multiplicities.
\end{proposition}
In particular, when the logical representation is irreducible, the multiplicity of the trivial representation in the decomposition of $\overline{\mathcal{C}}^*\otimes\mathcal{C}^*$ is equal to one. We refer to \cref{app:ProofPropo2} for a proof. One important implication of \cref{prop:lambdaRep} is that in order to be able to protect against physical errors that transform trivially, the logical representation must be irreducible.

\subsection{Choice of initial coherent state}\label{sec:InitState}
From \cref{Eq:CovEnc} one understands that codewords are built from orbits of an initial physical state (chosen at step (4) from~\cref{subsec:encoding}) under the action of $G$. The choice of initial physical state directly affect the fidelity of the code. The orbit construction is the quantum counterpart of classical group codes, in which a constellation of signals is generated as the orbit
\begin{align}
    G\vec{x}_0 = \{g\vec{x}_0 \: | \: g\in G\}
\end{align}
where $\vec{x}_0$ is an initial vector and the code is defined as the orbit of the state under the action of $G$~\cite{Slepian1968,ericson2001}. One can define a minimal distance between the initial vector $\vec{x}_0$ and vectors in the group orbit $g\vec{x}_0$ as,
\begin{align}
    d_{\mathrm{min}} &= \min_{a\not\in S}\Vert ax_0-x_0\Vert
\end{align}
where $S$ is the stabilizer subgroup of the initial vector and the norm is chosen such that the inner product is G-invariant. Because the noise vector under the additive white noise gaussian channel classically considered is spherically symmetric in signal space, no direction is preferred and performance is governed by the minimal Euclidean distance. The optimal constellation is often the one that spreads codewords as evenly as possible, therefore maximizing $d_{\mathrm{min}}$ is desired~\cite{Shannon1949}. 

It is natural to carry the reasoning to the quantum setting. For quantum codes, one can define an equivalent distance $d_{\mathrm{min}}^Q$ defined as,
\begin{align}
    d_{\mathrm{min}}^Q &= \min_{g\not\in S'}\Vert U_g \ket{\psi_0}-\ket{\psi_0}\Vert \label{Eq:TammesQuantum}
\end{align}
where the $G$-invariance of the inner product is automatic since we restrict to unitary representations. By analogy with classical group codes, one would expect that maximizing $d^{Q}_{\min}$ would lead to optimal choices of initial state. 

However, from numerical calculations, we conclude that this assumption does not hold for our code constructions~\cite{Ahmed2026}. In the following codes, we optimize the near-optimal fidelity of \cref{NearOptFid} over the initial two-mode coherent state $\ket{\sqrt{r}\alpha,\sqrt{1-r}e^{i\theta}\alpha}$ against the amplitude damping channel, where $r$ is the energy ratio between the first mode and the total energy and $e^{i\theta}$ is the relative phase between modes. What is found is that the optimal initial state does not yield a uniformly distributed orbit. We attribute this to the fact that the photon loss channel does not treat the 2N-dimensional phase space isotropically, whereas minimizing \cref{Eq:TammesQuantum} does not distinguish between directions in phase space.

Furthermore, for a fixed photon loss parameter $\gamma$ we observe an optimal size $|\alpha|$ that maximizes the fidelity, a situation similar to single-mode cat codes. On the one side, increasing $|\alpha|$ allows to better distinguish the different coherent states of the constellation. On the other side, increasing $|\alpha|$ also increases the probability of photon loss, and as a result the probability of an uncorrectable error. There is thus an optimal $|\alpha|$ that balances between these two requirements. 

For a qualitative exposition of the discrepancy between the optimization over the two metric, we refer to \cref{app:NearOptFidOptimization} where we present a side-by-side comparison of $\Tilde{F}$ and $d_{\mathrm{min}}^Q$ as a function of $r$ and $\theta$. We show that the optimization does not result in the same code and that the general symmetries of the two plot do not coincide. 
\section{Single-mode codes}\label{sect:singlemodecode}
In this section we analyze single-mode covariant codes based on initial coherent states. The aim of this section is to present pedagogical examples of the theory introduced in the previous section.  In particular, we follow the steps presented in \cref{sec:generalConstruction}, then compute the basic error-correcting properties of these codes using representation theory. All codes in this section correspond to a member of the so-called cat code family~\cite{Mirrahimi2014}, and much of the results presented here can be found in other forms elsewhere, see for example Ref.~\cite{Albert2018}.

\subsection{Four-legged cat code}
This code was proposed in~\cite{Mirrahimi2014} and realized in the milestone experiment of~\cite{Ofek2016}.
\subsubsection{Encoding description}
\begin{enumerate}
    \item $G = C_4 = \langle x \mid x^4=e \rangle$ and its character table is shown in \cref{tab:CharacterC4}. The cyclic group of order four has four one-dimensional irreducible representations labeled by $m \in [1,4]$, defined by: $\rho_m(x) = i^{m-1}$. The trivial representation corresponds to $m=1$.

    \item The standard choice is $\rho_f = \rho_2$. This leads to a physical representation $\rho(x) = \exp(i\pi \had \ha/2)$.
    \item Here there are multiple choices of logical encodings, with the standard choice being $\lambda \cong \rho_1 \oplus \rho_3$, or in other words
    \begin{equation}
        \lambda(x) = \begin{pmatrix}
            1 & 0\\
            0 & -1
        \end{pmatrix} = Z.
    \end{equation}
    The kernel of that representation is $C_2 = \langle x^2 \rangle$, leading to a stabilizer group $S = \langle \exp(i\pi \had \ha) \rangle$.
    \item The phase of the coherent state does not affect the error-correcting properties and choosing $\alpha \in \mathbb R_+$ is standard, with an optimal $|\alpha|$ that depends on $\gamma$.
\end{enumerate}

\subsubsection{Error-correction properties}
Here $\mathrm{Sesq}(\mathcal C) \cong (\rho_1 \oplus \rho_3)\otimes (\rho_1 \oplus \rho_3) \cong \rho_1^2 \oplus \rho_3^2$. The first trivial representation is, as is always the case, carried by the (logical) identity matrix $I$. The group action of $G$ on $\bar {\mathcal C}^* \otimes \mathcal C^*$ is given by conjugation and the only non-trivial element is the $Z$ gate. As a result the other trivial representation is carried by $\Zcode$ and the two sign representations are given by $\Xcode$ and $\Ycode$.

Analyzing the effect of a single photon loss, we obtain $\rho(x ^j) \ha \rho(x^j)^\dag = (-i)^j\ha $, such that $\ha$ transforms as the irreducible representation $\rho_4$. The results above therefore show that the block $Q_{0,1}$ of the QEC matrix is zero since $\rho_4$ does not appear in $\bar {\mathcal C}^* \otimes \mathcal C^*$.

In the event of two photons lost, we obtain $\rho(x^j) \ha^2 \rho(x^j)^\dag = (-1)^j\ha^2 $, such that $\ha^2$ transforms as the irreducible representation $\rho_2$. There is therefore no intrinsic protection against the loss of two photons, and carrying out the calculation indeed shows that this error leads to a logical bit flip $\Xcode$.

Note that the choice of logical encoding $\lambda \cong \rho_2 \oplus \rho_4$ leads to similar error-correction capabilities. In fact, one can choose to redefine the code space after observing a change in photon parity rather than actively correct errors~\cite{Ofek2016}. With this choice of encoding, we note that $\mathrm{ker}[\lambda]$ is trivial. It is therefore important to consider instead the stabilizer group $S = \mathrm{ker}[\mathbb P(\lambda)]$.

\subsection{Cyclic codes}\label{subsec:cyclicCodes}
The analysis of the four-legged cat code generalizes to general cyclic groups $C_n$. Here we highlight one more instructive example of code introduced in Ref.~\cite{Wetherbee2026}. We lay out its construction for the specific group $C_8$.
\subsubsection{Encoding description}
\begin{enumerate}
    \item $G = C_8 = \langle x \mid x^8=e\rangle$. There are eight possible one-dimensional representations of that group defined by $\rho_m(x) =\omega^{m-1}$, where $\omega= \exp(i \pi/4)$ and $m \in [1,8]$. Its character table is shown in \cref{tab:CharacterC8}. 
    \item The standard choice is $\rho_f = \rho_2$. This leads to a physical representation $\rho(x) = \exp(i\pi \had \ha/4)$
    \item The standard choice of encoding for a rotation-symmetric code would be $\lambda \cong \rho_1 \oplus \rho_5$, but another interesting choice is $\lambda \cong \rho_1 \oplus \rho_4$, or in other words
    \begin{equation}
        \lambda(x) = \begin{pmatrix}
            1 & 0\\
            0 & \omega^3
        \end{pmatrix} = T^3,
    \end{equation}
    where $T$ is a logical $T$ gate.
    \item The standard choice is $\alpha \in \mathbb R_+$, without effect on the error-correcting properties of the code.
\end{enumerate}

\subsubsection{Error-correction properties}
The kernel of the logical representation is trivial, $\mathrm{ker}[\mathbb P (\lambda)] = \{e\}$, and therefore there are no non-trivial stabilizers. However, this \emph{does not} imply that the code cannot correct errors.

For this choice of logical representation, we get $\mathrm{Sesq}(\mathcal C) \cong (\rho_1^* \oplus \rho_4^*)\otimes (\rho_1 \oplus \rho_4) \cong \rho_1^2 \oplus \rho_4 \oplus \rho_6$. The two trivial representations are carried by $\Icode$ and $\Zcode$. The representation $\rho_6$ is carried by $\Xcode + i \Ycode$ and the representation $\rho_4$ is carried by $\Xcode - i \Ycode$

Analyzing the effects of photon loss, we obtain that $\ha$ and $\ha^2$ transforms as the irreducible representations $\rho_8$ and $\rho_7$, respectively. Therefore, the blocks $Q_{0,1}$ and $Q_{0,2}$ of the QEC matrix are 0 since $\rho_7$ and $\rho_8$ do not appear in $\bar {\mathcal C}^* \otimes \mathcal C^*$. While this code does not have stabilizers, it can detect up to two lost photons.

In the event of three photons lost, we obtain that $\ha^3$ transforms as $\rho_6$. There is a single irreducible representation $\rho_6$ in $\bar {\mathcal C}^* \otimes \mathcal C^*$, which correspond to $\Xcode + i \Ycode$. Without explicitly computing the QEC matrix element $Q_{0,3}$, we know that it is proportional to a logical decay event.
\section{Two-mode codes}\label{sect:twomodecode} 
Having defined a single mode codes, we analyze new and instructive codes within our construction based on two bosonic modes and non-cyclic groups. More details on two-mode covariant codes based on cyclic groups can be found in Ref.~\cite{Ahmed2026}.
In particular, we show that the construction of the previous section allows to extract error-correction properties in a straightforward manner.

\subsection{Quaternion Codes}
The quaternion codes constitute one of the simplest non-trivial cases of two-mode bosonic codes within the proposed construction. The quaternion group $Q_8$ is an example of a binary dihedral group isomorphic to the dicyclic group $\mathrm{Dic}_2$ when considered as the double cover of the dihedral group $D_4$. The quaternion group has a presentation given by $Q_8 = \langle -e,i,j,k \mid (-e)^2=e, i^2=j^2=k^2=ijk=-e\rangle$ and its character table is shown in \cref{tab:CharacterQ8}. The only two-dimensional irreducible representation $\rho_5$ is given by, 
\begin{align}
    \rho_5(i) = \begin{bmatrix}
        i&0\\0&-i
    \end{bmatrix}, &&
    \rho_5(j) = \begin{bmatrix}
        0&-1\\1&0
    \end{bmatrix}, \nonumber \\
    \rho_5(-e) = \begin{bmatrix}
        -1&0\\0&-1
    \end{bmatrix}, && \rho_5(k) = \begin{bmatrix}
        0&-i\\-i&0
    \end{bmatrix}. \nonumber
\end{align}
\begin{table}[H]
\caption{%
Choices of logical representation for the $Q_8$ qubit}\label{tab:table1}
\begin{ruledtabular}\label{table:Q8LogicalReps}
\begin{tabular}{ccc}
\makecell[t]{Logical\\representation} &
\makecell[t]{Isomorphism class of\\ the stabilizer group} &
\makecell[t]{Logical gates\\group} \\
\colrule
$\rho_1\oplus\rho_1$ & $Q_8$ & $\{e\}$ \\
$\rho_1\oplus\rho_2$ & $C_4$ & $C_2$  \\
$\rho_1\oplus\rho_3$ & $C_4$ & $C_2$  \\
$\rho_1\oplus\rho_4$ & $C_4$ & $C_2$  \\
$\rho_2\oplus\rho_2$ & $Q_8$ & $\{e\}$  \\
$\rho_2\oplus\rho_3$ & $C_4$ & $C_2$  \\
$\rho_2\oplus\rho_4$ & $C_4$ & $C_2$  \\
$\rho_3\oplus\rho_3$ & $Q_8$ & $\{e\}$  \\
$\rho_3\oplus\rho_4$ & $C_4$ & $C_2$  \\
$\rho_4\oplus\rho_4$ & $Q_8$ & $\{e\}$  \\
$\rho_5$ & $C_2$ & $V_4$  \\
\end{tabular}
\end{ruledtabular}
\end{table}
Several choices of logical representations are listed in \cref{table:Q8LogicalReps}. 
As an example of this construction, we choose the logical representation $\lambda=\rho_2\oplus\rho_4$ with stabilizer group isomorphic to $\mathrm{ker}[\mathbb{P}(\rho_2\oplus\rho_4)]=C_4$. 

\begin{figure*}
    \centering
    \includegraphics[width=\textwidth]{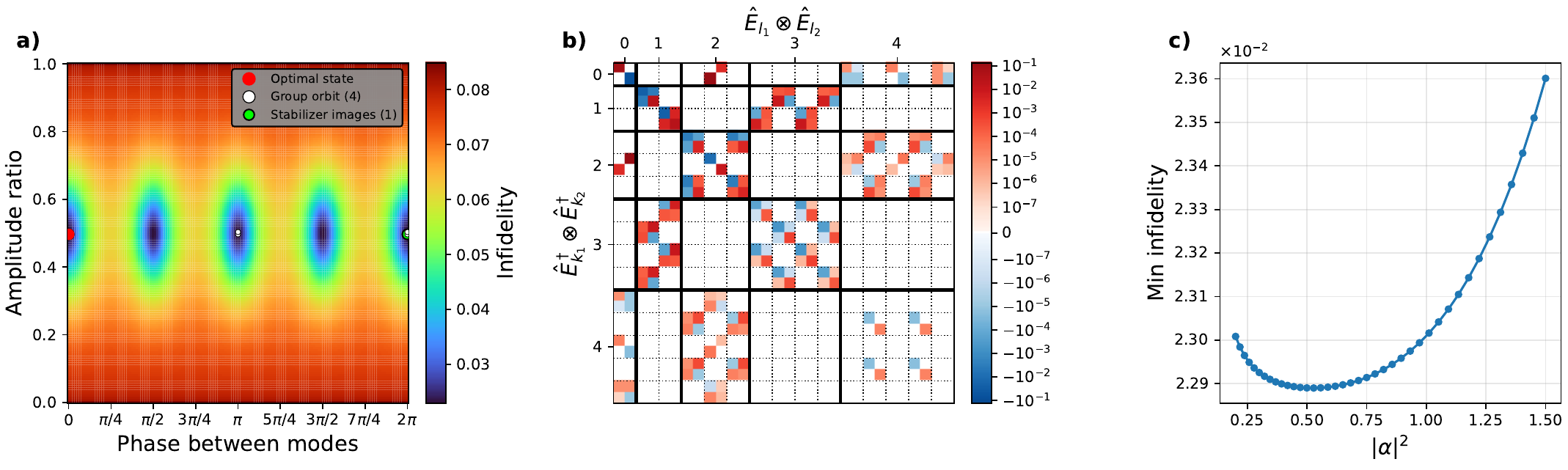}
    \caption{Error-correction performance of the two-mode quaternion ($Q_8$) code with logical representation $\lambda\cong \rho_2\oplus\rho_4$, evaluated against the amplitude damping channel for $\gamma = 0.095$. a) Near-optimal infidelity, $1-\tilde F$,~\cref{NearOptFid} as a function of the initial coherent state $\ket{\sqrt{r}\alpha,\sqrt{1-r}e^{i\theta}\alpha}$ where $r$ is the energy ratio between the first mode and the total energy and $\theta$ is the relative phase between modes. The near-optimal fidelity is plotted for the optimal initial state energy and results in $1-\tilde F\approx 0.023$. b) Uncorrectable part of the QEC matrix plotted for the optimal initial parameters $r=0.5$, $\theta=0$ and $|\alpha|^2\approx0.528$. Rows and columns are indexed by the two-mode loss label $\hat{E}_{k_1}^\dagger\otimes\hat{E}_{k_2}^\dagger$ and $\hat{E}_{l_1}\otimes\hat{E}_{l_2}$ and grouped (solid lines) by total error order $|l| = l_1 + l_2$. c) Minimum near-optimal infidelity optimized over $r$ and $\theta$ as a function of total energy $|\alpha|^2$.}
    \label{fig:Q8-24}
\end{figure*}

The logical group is then given by $L=Q_8/\mathrm{ker[\mathbb{P}(\lambda)]}\cong C_2$ and the two cosets are,
\begin{equation}
\begin{aligned}
\lambda(\{\pm e, \pm i\}) &=
\begin{bmatrix}
    1&0\\ 0 & 1
    \end{bmatrix},\\
    \lambda(\{\pm j, \pm k\}) &= \begin{bmatrix}
        1&0\\0&-1
    \end{bmatrix},
    \end{aligned}
\end{equation}
such that the only nontrivial logical is a Z gate.
Using the projection formula given by \cref{Projection} the codewords are 
\begin{align}
    \ket{0} &= \mathcal{N}_0 \sum_{Q_8}\chi_{\rho_2}^*(g)\rho_5(g)\ket{\alpha,\beta} \nonumber \\
    &= \mathcal{N}_0 [\ket{\alpha,\beta}+\ket{-\alpha,-\beta}-\ket{i\alpha,-i\beta}-\ket{-i\alpha,i\beta}\nonumber \\
    &+\ket{-\beta,\alpha}+\ket{\beta,-\alpha}-\ket{-i\beta,-i\alpha}-\ket{i\beta,i\alpha}]\\
    \ket{1} &= \mathcal{N}_1 \sum_{Q_8}\chi_{\rho_4}^*(g)\rho_5(g)\ket{\alpha,\beta} \nonumber \\
    &= \mathcal{N}_1 [\ket{\alpha,\beta}+\ket{-\alpha,-\beta}-\ket{i\alpha,-i\beta}-\ket{-i\alpha,i\beta}\nonumber \\
    &-\ket{-\beta,\alpha}-\ket{\beta,-\alpha}+\ket{-i\beta,-i\alpha}+\ket{i\beta,i\alpha}]
\end{align}
where $\mathcal{N}_0,\mathcal{N}_1$ are normalization constants. Here $\mathrm{Sesq}(\mathcal C) \cong (\rho_2 \oplus \rho_4)\otimes (\rho_2 \oplus \rho_4) \cong \rho_1^2 \oplus \rho_3^2$. The two trivial representations in $\mathrm{Sesq}(\mathcal C)$ are given by $I$ and $Z$. The two representations $\rho_3$ are given by $X$ and $Y$.
Since error operators transform under conjugation like the physical representation $\mathrm{Sym}^{L}(\rho_5^*)$, we can predict the effect of photon loss on the logical subspace. The results are consolidated in \cref{tab:DecompoQ8}. 
\begin{table}[H]
\caption{
Decomposition $\mathrm{Sym}^{K}(\rho_5)\otimes \mathrm{Sym}^{L}(\rho_5^*)$ and logical errors for $\lambda\cong\rho_2\oplus\rho_4$ of group $Q_8$
}\label{tab:DecompoQ8}%
\begin{ruledtabular}
\begin{tabular}{c c c}
\textrm{$(K,L)$} &
\parbox[c]{0.32\columnwidth}{\centering
$\mathrm{Sym}^{K}(\rho_5)\otimes \mathrm{Sym}^{L}(\rho_5^*)$
} &
\parbox[c]{0.38\columnwidth}{\centering
Logical Errors
} \\
\colrule
$(0,0)$ & $\rho_1$ & $\Zcode$ \\
$(0,1)$ & $\rho_5$ & - \\
$(0,2)$ & $\rho_2\oplus\rho_3\oplus\rho_4$ & $\Xcode,\Ycode$ \\
$(1,1)$ & $\rho_1\oplus\rho_2\oplus\rho_3\oplus\rho_4$ & $\Zcode,\Xcode,\Ycode$\\
\end{tabular}
\end{ruledtabular}
\end{table}
The single photon loss errors transform as $\rho_5$ and the $Q_{0,1}$ block of the QEC matrix is zero, as confirmed in \cref{fig:Q8-24}b). The two photon loss error sector contains the isotypic component of the codespace suggesting that correction is not possible for two photon losses. A further decomposition shows that the $\rho_3$ irreducible representation in the $(0,2)$ block is carried by the error $\hat a \hat b$, which is precisely the undetectable error shown in \cref{fig:Q8-24}b). In that same block, the errors $\hat a^2 \pm \hat b^2$ carry the representations $\rho_2$ and $\rho_4$, respectively, such that the corresponding QEC elements are zero. 

\Cref{fig:Q8-24}b,c) were computed using the initial state $\ket{\alpha/\sqrt2,\alpha/\sqrt2}$, but panel a) shows that an equally good choice is $\ket{\alpha/\sqrt2,i\alpha/\sqrt2}$, which yields a different code. Choosing the initial state $\ket{\alpha/\sqrt2,-\alpha/\sqrt2}$ is completely equivalent to choosing $\ket{\alpha/\sqrt2,\alpha/\sqrt2}$ with a different logical basis. Panel c) shows that, as with other cat codes, there is an optimal $|\alpha|$ that balances between dephasing due to coherent contraction and number of induced uncorrectable errors. 

\subsection{Binary octahedral code}\label{sec:BinaryO}
The binary octahedral group ($2O$) is a finite group of order 48. The group has a presentation given by $2O = \langle x,y,z \mid x^2=y^4=z^3=xzy=-e\rangle$; its character table appears in \cref{tab:Character2O}. The group $2O$ has two two-dimensional faithful irreducible representations $\rho_4$ and $\rho_5$. We choose as a physical representation $\rho_4$. In order to choose a stabilizer subgroup, we look at the kernel of all two-dimensional representations. \Cref{tab:Logical2O} shows all possible choices.
\begin{table}[H]
\caption{
Choices of logical representation for the $2O$ qubit}\label{tab:Logical2O}
\begin{ruledtabular}
\begin{tabular}{ccc}
\makecell[t]{Logical\\representation} &
\makecell[t]{Isomorphism class of\\ the stabilizer group} &
\makecell[t]{Logical gates\\group} \\
\colrule
$\rho_1\oplus\rho_1$ & $2O$ & $\{e\}$ \\
$\rho_1\oplus\rho_2$ & $2T$ & $C_2$ \\
$\rho_2\oplus\rho_2$ & $2O$ & $\{e\}$ \\
$\rho_3$ & $Q_8$ & $S_3$ \\
$\rho_4$ & $C_2$ & $S_4$ \\
$\rho_5$ & $C_2$ & $S_4$ \\
\end{tabular}
\end{ruledtabular}
\end{table}
In the $\rho_4$ representation, the generators of the group $2O$ are given by,
\begin{align}
    \rho_4(x) &= \frac{1}{\sqrt{2}}\begin{bmatrix}i&1\\-1&-i\end{bmatrix},\\
    \rho_4(y) &= \frac{1}{\sqrt{2}}\begin{bmatrix}
        1+i&0\\0&1-i\end{bmatrix},\\
    \rho_4(z) &= \frac{1}{2}\begin{bmatrix}
        1+i&1+i\\-1+i&1-i
    \end{bmatrix}.
\end{align}
We proceed here in building two detailed example of codes with stabilizer group isomorphic to $2T$ and $Q_8$ respectively.
\subsubsection{Stabilizer group isomorphic to $2T$}
\begin{figure*}[tp]
    \centering
    \includegraphics[width=\textwidth]{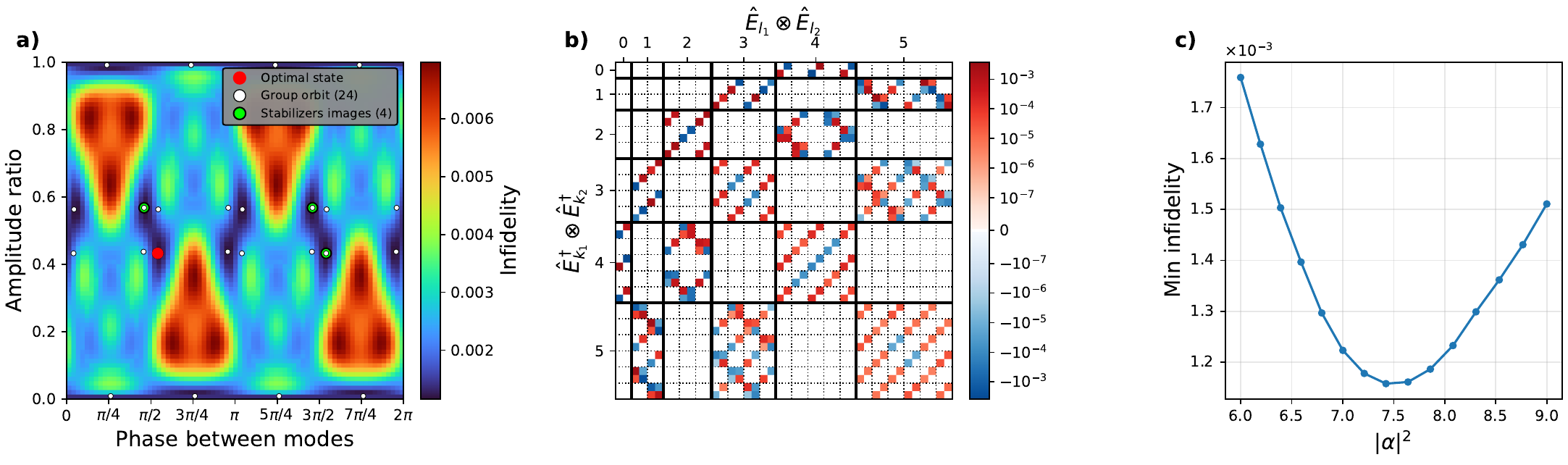}
    \caption{Error-correction performance of the two-mode binary octahedral group ($2O$) code with logical representation $\lambda\cong \rho_1\oplus\rho_2$, evaluated against the amplitude damping channel. a) Near-optimal infidelity, $1-\tilde F$,~\cref{NearOptFid} as a function of the initial coherent state $\ket{\sqrt{r}\alpha,\sqrt{1-r}e^{i\theta}\alpha}$ where $r$ is the energy ratio between the first mode and the total energy and $\theta$ is the relative phase between modes. The near-optimal fidelity is plotted for the optimal initial state energy and results in $1-\tilde F\approx 0.0009$. b) Uncorrectable part of the QEC matrix for loss-rate $\gamma=0.095$ plotted for the optimal initial parameters $r\approx0.47$, $\theta\approx1.33\;\mathrm{rad}$ and $|\alpha|^2\approx6.31$. Rows and columns are indexed by the two-mode loss label $\hat{E}_{k_1}^\dagger\otimes\hat{E}_{k_2}^\dagger$ and $\hat{E}_{l_1}\otimes\hat{E}_{l_2}$ and grouped (solid lines) by total error order $|l| = l_1 + l_2$. c) Minimum near-optimal infidelity optimized over $r$ and $\theta$ as a function of total energy $|\alpha|^2$.}
    \label{fig:2O-12}
\end{figure*}

Choosing the logical representation to be $\lambda \cong \rho_1\oplus\rho_2$, the stabilizer group is given by $\mathrm{ker}[\mathbb{P}(\rho_1\oplus\rho_2)]\cong 2T$ and the logical group is $L=2O/\mathrm{ker}[\mathbb{P}(\lambda)]\cong 2O/2T\cong C_2$. The codewords are built using the projection formula of \cref{Projection}. 
\begin{align}
    \ket{0} &= \mathcal{N}_0 \sum_{2O}\rho_5 (g)\ket{\alpha,\beta}\\
    \ket{1} &= \mathcal{N}_1 \sum_{2O}\chi^*_{\rho_2}(g)\rho_5(g)\ket{\alpha,\beta}
\end{align}
where $\mathcal{N}_0,\mathcal{N}_1$ are normalization constants. Here $\mathrm{Sesq}(\mathcal C) \cong (\rho_1 \oplus \rho_2)\otimes (\rho_1 \oplus \rho_2) \cong \rho_1^2 \oplus \rho_2^2$, with the representations $\rho_2$ carried by $X$ and $Z$. This code protects against loss of three photons, as understood from \cref{tab:2O-12} and confirmed from the QEC matrix of \cref{fig:2O-12}b). One interesting feature of this code is that $\rho_2$ only appears in the decomposition of the blocks $(K,L)$ where $K+L \geq 6$, yielding a logical qubit where $\Xcode$ or $\Ycode$ logical errors only appear at order 6 in $\gamma$.
\begin{table}[H]
\caption{\label{tab:2O-12}%
Decomposition of $\mathrm{Sym}^{K}(\rho_4)\otimes \mathrm{Sym}^{L}(\rho_4^*)$ and logical errors for $\lambda\cong\rho_1\oplus\rho_2$ of the group $2O$
}
\begin{ruledtabular}
\begin{tabular}{c c c}
\textrm{$(K,L)$} &
\parbox[c]{0.32\columnwidth}{\centering
$\mathrm{Sym}^{K}(\rho_4)\otimes \mathrm{Sym}^{L}(\rho_4^*)$
} &
\parbox[c]{0.38\columnwidth}{\centering
Logical Errors
}\\
\colrule
$(0,0)$ & $\rho_1$ & $\Zcode$ \\
$(0,1)$ & $\rho_4$ & - \\
$(0,2)$ & $\rho_6$ & - \\
$(1,1)$ & $\rho_1\oplus\rho_6$ & $\Zcode$ \\
$(0,3)$ & $\rho_8$ & -\\
$(2,2)$ & $\rho_1\oplus\rho_3\oplus\rho_6\oplus\rho_7$ & $\Zcode$ \\
$(3,3)$ & $\rho_1\oplus\rho_2\oplus\rho_3\oplus\rho_6^2\oplus\rho_7^2$ & $\Zcode, \Xcode, \Ycode$\\
\end{tabular}
\end{ruledtabular}
\end{table}
\subsubsection{Stabilizer group isomorphic to $Q_8$}
\begin{figure*}[ht]
    \centering
    \includegraphics[width=\textwidth]{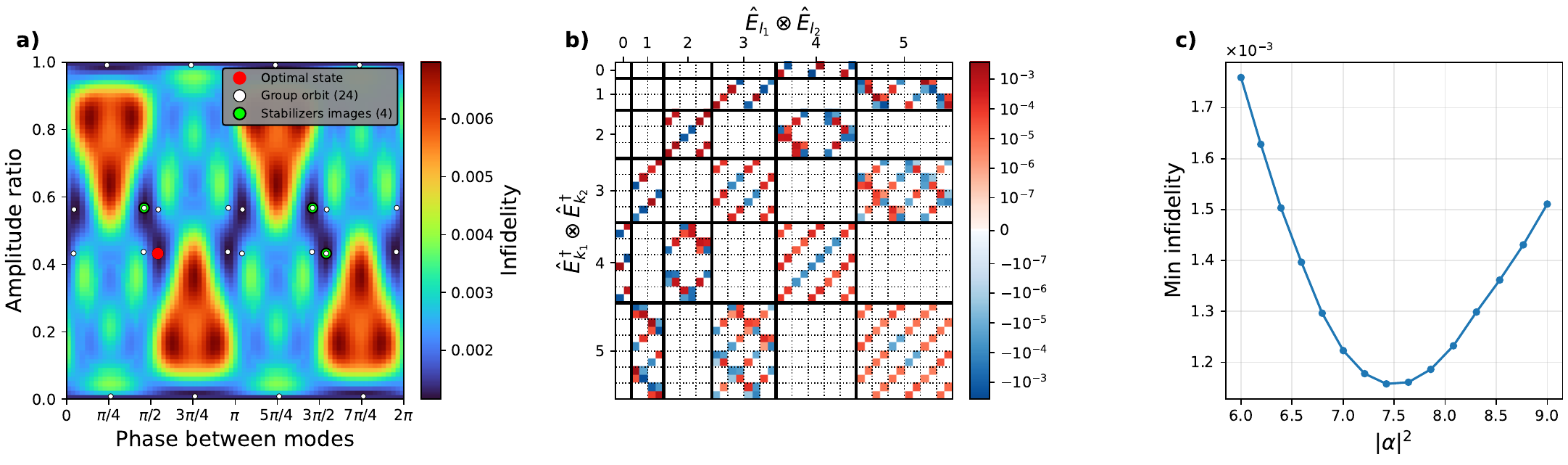}
    \caption{Error-correction performance of the two-mode binary octahedral group ($2O$) code with logical representation $\lambda\cong \rho_3$, evaluated against the amplitude damping channel. a) Near-optimal infidelity, $1-\tilde F$,~\cref{NearOptFid} as a function of the initial coherent state $\ket{\sqrt{r}\alpha,\sqrt{1-r}e^{i\theta}\alpha}$ where $r$ is the energy ratio between the first mode and the total energy and $\theta$ is the relative phase between modes. The near-optimal fidelity is plotted for the optimal initial state energy and results in $1-\tilde F\approx 0.0012$. b) Uncorrectable part of the QEC matrix for loss-rate $\gamma=0.095$ plotted for the optimal initial parameters $r\approx0.43$, $\theta\approx1.7\;\mathrm{rad}$ and $|\alpha|^2\approx7.42$. Rows and columns are indexed by the two-mode loss label $\hat{E}_{k_1}^\dagger\otimes\hat{E}_{k_2}^\dagger$ and $\hat{E}_{l_1}\otimes\hat{E}_{l_2}$ and grouped (solid lines) by total error order $|l| = l_1 + l_2$. c) Minimum near-optimal infidelity optimized over $r$ and $\theta$ as a function of total energy $|\alpha|^2$.}
    \label{fig:2O-3}
\end{figure*}

Starting with the abstract group $G = 2O$, but choosing instead a two-dimensional logical irreducible representation, $\lambda \cong \rho_3$, we obtain an example of code with a nontrivial set of gates and a non-abelian stabilizer group, $\mathrm{ker}[\mathbb{P}(\rho_3)]\cong Q_8$. The logical group is then isomorphic to the symmetric group $S_3$. The logical gates are given explicitly by the two following generators
\begin{align}
    \lambda\left(\frac{1+i+j+k}{\sqrt{2}}\right) &= \begin{bmatrix}
        \omega&0\\0&\omega^2
    \end{bmatrix},\\
    \lambda\left(\frac{1+i}{\sqrt{2}}\right) &= \begin{bmatrix}
        0&1\\1&0
    \end{bmatrix},
\end{align}
with $\omega = e^{2\pi i/3}$. With this choice, $\mathrm{Sesq}(\mathcal C) \cong \rho_3 \otimes \rho_3 \cong \rho_1 \oplus \rho_2 \oplus \rho_3$, with a single trivial representation present. The trivial representation is carried by $\Icode$, $\rho_2$ corresponds to $\Zcode$ and the two-dimensional representation $\rho_3$ correspond to $\mathrm{span}\{\Xcode,\Ycode\}$. Unlike codes constructed until now, the choice of this logical representation perfectly protects against the ``no-jump'' part of the photon loss Kraus operators, as understood from \cref{tab:2O-3} and the QEC matrix of \cref{fig:2O-3}b).
\begin{table}[H]
\caption{\label{tab:2O-3}%
Decomposition of $\mathrm{Sym}^{K}(\rho_4)\otimes \mathrm{Sym}^{L}(\rho_4^*)$ and logical errors for $\lambda\cong\rho_3$ of the group $2O$
}
\begin{ruledtabular}
\begin{tabular}{c c c}
\textrm{$(K,L)$} &
\parbox[c]{0.32\columnwidth}{\centering
$\mathrm{Sym}^{K}(\rho_4)\otimes \mathrm{Sym}^{L}(\rho_4^*)$
} &
\parbox[c]{0.38\columnwidth}{\centering
Logical Errors
}\\
\colrule
$(0,0)$ & $\rho_1$ & - \\
$(0,1)$ & $\rho_4$ & - \\
$(0,2)$ & $\rho_6$ & - \\
$(1,1)$ & $\rho_1\oplus\rho_6$ & - \\
$(0,3)$ & $\rho_8$ & - \\
$(2,2)$ & $\rho_1\oplus\rho_3\oplus\rho_6\oplus\rho_7$ & $\Xcode,\Ycode$ \\
$(3,3)$ & $\rho_1\oplus\rho_2\oplus\rho_3\oplus\rho_6^2\oplus\rho_7^2$ & $\Zcode, \Xcode, \Ycode$\\
\end{tabular}
\end{ruledtabular}
\end{table}

\subsection{Other known codes}
In this subsection, we make contact with other two-mode cat codes found in the literature that can be expressed using the covariant construction. As we have done above, we list the series of choices laid out in \cref{subsec:encoding} that lead to equivalent codes.
\subsubsection{Pair-cat code}
While we focused in this work on finite subgroups of SU(2), the covariant encoding approach also extends to infinite (compact) groups~\cite{Faist2020,Zhou2021,Liu2021}. One example is the pair-cat code introduced in Ref.~\cite{Albert2019}, which can be reframed within the current construction. 
\begin{enumerate}
    \item $G = O(2)$, a group defined by continuous rotations $R(\theta), \theta \in [0,2\pi)$ and a reflection $r$ such that $r^2 = e, r R(\theta) r = R(-\theta)$. There are two one-dimensional representations of this group given by the trivial representation $\rho_{\mathrm{triv}}$ and the sign representation defined by $\rho_{\mathrm{sign}}(R(\theta)) = 1, \rho_{\mathrm{sign}}(r) = -1$. There are an infinite number of two-dimensional representations indexed by $k \in \mathbb N^*$, for example defined by 
    \begin{equation}
        \rho_k(r) = \begin{pmatrix}
            0 & 1 \\
            1 & 0
        \end{pmatrix};\quad \rho_k[R(\theta)] = \begin{pmatrix}
            e^{i k \theta} & 0\\
            0 & e^{-i k \theta}
        \end{pmatrix}
    \end{equation}
    \item The natural choice is $\rho_f = \rho_1$.
    \item The choice made in Ref.~\cite{Albert2019} (up to a change of logical basis) is $\lambda \cong \rho_{\mathrm{triv}} \oplus \rho_\mathrm{sign}$, with stabilizer group $\mathrm{ker}[\mathbb P (\lambda)]\cong U(1)$ given by the continuous rotations.
    \item The choice made in Ref.~\cite{Albert2019} is simply $\ket{\alpha,\alpha}$.
\end{enumerate}
Since the codewords are eigenstates of $R(\theta)$ for all values of $\theta$, that means that the codewords are also eigenstates of the generator for $\rho(R(\theta))$, i.e. the photon number difference operator $\hat a^\dag \hat a - \hat b^\dag \hat b$.

\subsubsection{Dual-rail cat codes}
Ref.~\cite{Biswas2026} introduced dual-rail cat codes, which can be reframed as follow:
\begin{enumerate}
    \item $G = D_4 = \langle x,z \mid x^2 = z^2=e,(xz)^4=e \rangle$. The character table for this code is identical to the group $Q_8$, see \cref{tab:CharacterQ8}, with $\rho_5$ being the only two-dimensional irreducible representation.
    \item The choice made in Ref.~\cite{Biswas2026} is
    \begin{equation}
        \rho_5(x) = \begin{pmatrix}
            0 & 1 \\
            1 & 0
        \end{pmatrix};\quad
        \rho_5(z) = \begin{pmatrix}
            -1 & 0 \\
            0 & 1
        \end{pmatrix}
    \end{equation}
    \item The $X$ and $Z$ logical gates can be realized using this code by choosing $\lambda \cong \rho_5$, with $\mathbb P(\lambda) \cong V_4$ and  $\mathrm{ker}[\mathbb P (\lambda)]\cong C_2$.
\end{enumerate}
Ref.~\cite{Biswas2026} also discussed dual-rail four-legged cat codes which can be obtained by starting from the group $G = D_8$.

\subsubsection{Cyclic simplex code}
In a similar way to the single-mode cyclic code~\cite{Wetherbee2026} discussed in \cref{subsec:cyclicCodes}, one can sometimes make small modifications to a code to add logical gates at a minimal cost in the number of correctable errors by choosing different logical representations. A two-mode example of this is the simplex code introduced in Ref.~\cite{Jain2024}.
\begin{enumerate}
    \item $G = \mathbb Z_5 \times \mathbb Z_2 \cong \mathbb Z_{10} = \langle x\mid x^{10}=e \rangle$. There are ten possible one-dimensional representations of that group defined by $\rho_m(x) = \omega^{m-1}$, where $\omega= \exp(i \pi/5)$ and $m \in [1,10]$. The character table is shown in \cref{tab:CharacterC10}.
    \item The choice made in Ref.~\cite{Jain2024} is the (reducible) representation given by
    \begin{equation}
        \rho(x) = \begin{pmatrix}
            \omega & 0 \\
            0 & \omega^2
        \end{pmatrix}.
    \end{equation}
    \item The choice made in Ref.~\cite{Jain2024} is $\lambda \cong \rho_1 \oplus \rho_6$, with $\mathrm{ker}[\mathbb P (\lambda)]\cong \mathbb Z_5$. Ref.~\cite{Wetherbee2026} pointed out that one can add logical gates to this code by choosing instead $\lambda \cong \rho_1 \oplus \rho_4$, a code that can still protect against two photon loss although $\mathrm{ker}[\mathbb P (\lambda)]\cong \{e\}$.
    \item Again $\ket{\alpha,\alpha}$ is a natural choice.
\end{enumerate}

\section{Conclusion}
We have expanded on the representation-theoretic framework of QEC for the construction and analysis of multimode covariant bosonic codes built from finite groups. The construction takes as input an abstract group $G$, a physical representation $\rho$ and a logical representation $\lambda$, from which codewords follow by projection or covariant encoding. Our central observation is that the space of QEC matrices is a morphism of group representations; then by Schur's lemma, its entries vanish whenever an error sector and the code space share no irreducible component. The Knill-Laflamme conditions therefore reduces to comparing the decompositions of $\mathrm{Sym}^L(\rho_f)\otimes\mathrm{Sym}^K(\rho_f^*)$ and $\overline{\mathcal{C}}^*\otimes \mathcal{C}^*$ into irreducible representations. Detectability and correctability of a given loss order reduce to character computations, removing the need of explicit calculation of the QEC matrix for several errors.

Within this construction we have recovered several known codes such as the four-legged cat code~\cite{Leghtas2013, Mirrahimi2014} and bosonic cyclic codes~\cite{Wetherbee2026}. More importantly, we obtained insights into new two-mode codes based on the quaternion group $Q_8$ and the binary octahedral group $2O$. The examples makes explicit the tradeoff in error protection governed by the reducibility of the logical representation. A reducible representation places the $\Zcode$ error in the same isotypic sector as the logical identity and therefore does not provide intrinsic protection against logical $\Zcode$ errors. On the other hand, an irreducible logical representation like in the $2O$ code with $\lambda\cong\rho_3$ protects against logical $\Zcode$ errors stemming from the ``no-jump'' part of the amplitude damping channel. The same representation-theoretic bookkeeping determines the stabilizer subgroup $\ker[\mathbb{P}(\lambda)]$ and the logical gate group $\mathbb{P}[\lambda(G)]$ once the global phase is removed. 

Finally, we optimized the initial state over total energy, energy ratio between the first mode and the total energy and relative phase between modes using the near-optimal fidelity as a figure of merit. Contrary to the intuition inherited from classical group code and quantum spherical code, the optimal initial state under this metric does not guarantee a uniformly distributed orbit. The near-optimal fidelity depends directly on the error channel of the system whereas the minimal distance $d_{\mathrm{min}}^Q$ does not.

Several directions could be explored within this framework. The construction extends naturally to qudits and finite subgroups of $U(N)$. It would be informative to characterize which groups and/or representations yield a balanced number of stabilizers and logical operations. 

Finally, the present analysis predicts what errors can be corrected in principle, but does not propose a systematic protocol to correct these errors in a practical context.

\begin{acknowledgments}
We thank Lucas Barbier and Lev-Arcady Sellem for useful conversations and proof reading of the manuscript. J.P.B. acknowledges the support of the Natural Sciences and Engineering Research Council of Canada (NSERC), funding reference number RGPIN-2020-05557. B.R. acknowledges the support of the Natural Sciences and Engineering Research Council of Canada (NSERC), funding reference number RGPIN-2022-04451, as well as the Fonds de Recherche Nature du Québec, Nature et Technologies (FRQNT).
\end{acknowledgments}

\appendix

\crefalias{section}{appendix}
\crefname{appendix}{App.}{Apps.}
\Crefname{appendix}{App.}{Apps.}

\section{Proof of proposition 1}\label{app:Proposition1}
Let $v\in\mathbb{C}^N$ be any vector and $w \in(\mathbb{C}^N)^*$ be any linear functional. The creation (annihilation) operator acts linearly (antilinearly) on $v$. The single-photon state is defined as $\ket{v}\coloneqq \hat{a}^\dagger(v)\ket{0} = \sum_j v_j \hat a^\dag_j \ket{0}$ and $\rho(g)$ acts on $\ket{v}$ via $\rho(g)\ket{v}=\ket{\rho_f(g)v}$. 

We define the following map
\begin{equation}
\begin{aligned}
    \Phi: (\mathbb{C}^N)^{*} &\longrightarrow \mathrm{Span}\{\hat{a}_1,\hat{a}_2,...,\hat{a}_N\}\\
    \Phi(w)\ket{v} &\coloneqq w(v)\ket{0}
\end{aligned}
\end{equation}
Eq.(25) can be rewritten as 
\begin{align}
    (\sigma(g)\Phi(w))\ket{v} &= \rho(g)\Phi(w)\rho^{-1}(g)\ket{v}\\
    &= w(\rho_f^{-1}(g)v)\rho(g)\ket{0}\\
    &= w(\rho_f^{-1}(g)v)\ket{0}\\
    &= (\rho_f^\dagger(g)w)(v)\ket{0}\\
    &= \Phi(\rho_f^\dagger(g)w)\ket{v}
\end{align}
One can see that $\sigma(g)\circ\Phi = \Phi\circ\rho_f^*(g),\:\forall g \in G$ and that indeed the map $\Phi$ is a morphism of representation between $\sigma$ and $\rho^*$. Therefore, annihilation operators transforms under the adjoint action of the representation $\rho$ like the representation $\sigma$ which is isomorphic to $\rho_f^*$.

\section{Proof of proposition 2}\label{app:ProofPropo3}
Given an error model and a subspace of operators $L$ where $L$ also needs to be a subrepresentation of $G$ over the physical error space, the map 
\[ 
    \Lambda : \overline{L}^*\otimes L^* \longrightarrow \mathrm{Sesq}(\mathcal{C})
\]
defined by $\Lambda(\hat{L}_{\boldsymbol{k}}\otimes\hat{L}_{\boldsymbol{l}})\coloneqq Q_{\boldsymbol{k},\boldsymbol{l}}$ is a morphism of representations.
Let $\mu,\nu\in \mathcal{H}_L$. Then,
\begin{equation}
\begin{aligned}
    \Lambda(\sigma(g)&(\hat{L}_{\boldsymbol{k}}\otimes\hat{L}_{\boldsymbol{l}}))(\ket{\mu},\ket{\nu})\\
    &= \Lambda(\sigma(g)\hat{L}_{\boldsymbol{k}}\otimes \sigma(g)\hat{L}_{\boldsymbol{l}})(\ket{\mu},\ket{\nu})\\
    &= \bra{\mu}V_G^\dagger\rho(g)\hat{L}_{\boldsymbol{k}}^\dagger \rho(g)^\dagger \rho(g)\hat{L}_{\boldsymbol{l}} \rho(g)^\dagger V_G\ket{\nu} \\
    &= \bra{\mu}\lambda(g)V_G^\dagger\hat{L}_{\boldsymbol{k}}^\dagger \hat{L}_{\boldsymbol{l}} V_G\lambda(g)^\dagger\ket{\nu}\\
    &= Q_{\boldsymbol{k},\boldsymbol{l}}(\lambda(g)^\dagger \ket{\mu}, \lambda(g)^\dagger \ket{\nu})\\
    &= g\star Q_{\boldsymbol{k},\boldsymbol{l}}(\ket{\mu}, \ket{\nu})
    \end{aligned}
\end{equation}

\section{Proof of proposition 3}\label{app:ProofPropo2}
Let $\lambda$ be a logical representation that is completely reducible and the conjugation action on logical operators gives $\mathrm{Sesq}(\mathcal{C})\cong \overline{\lambda}^*\otimes \lambda^*$. The multiplicity of the trivial representation in the decomposition of $\mathrm{Sesq}(\mathcal{C})$ into irreducible factors is given by $\langle \chi_\lambda,\chi_\lambda \rangle$.

The character of $\lambda$ is given by $\chi_\lambda = (c_1, c_2, ..., c_n)$. One can always write 
\begin{align}
    \lambda = \bigoplus_{i=1}^n\rho_i^{m_i},
\end{align}
where $n_i$ is the multiplicity of the irreducible representation $\rho_i$. The inner product of $\chi_\lambda$ with itself is given by
\begin{align}
    \langle \chi_\lambda, \chi_\lambda \rangle &= \frac{1}{|G|}\sum_G \overline{\chi_\lambda}(g)\chi_\lambda(g)\label{eq:NormChiLambda}
\end{align}
The character of $\chi_{\lambda\otimes\lambda^*}=\chi_\lambda \chi_{\lambda^*}=(|c_1|^2, |c_2|^2, ..., |c_n|^2)$. To find the multiplicity of the trivial representation in the decomposition of $\lambda\otimes\lambda^*$, we calculate 
\begin{equation}
    \langle\chi_{\rho_1}, \chi_{\lambda\otimes\lambda^*}\rangle=1/|G|\sum_g \overline{\chi_\lambda}(g)\chi_\lambda(g) \label{eq:DecompLambdaLambda}.
\end{equation}
We see that \cref{eq:NormChiLambda} and \cref{eq:DecompLambdaLambda} coincide. Hence, the multiplicity of the trivial irreducible representation in the decomposition of $\lambda\otimes\lambda^*$ is given by the inner product $\langle \chi_\lambda, \chi_\lambda \rangle = \sum_i m_i^2$.  Finally, for a qubit ($d=2$), the logical Paulis $\{\Icode, \Xcode, \Ycode, \Zcode\}$ span $\mathrm{Sesq}(\mathcal{C})$ and $\Icode$ is invariant under conjugation so it always carries a copy of the trivial representation. In the qubit case, given a choice of basis, when $\lambda$ is reducible, another trivial copy is carried by $\Zcode$ forcing $\Icode$ and $\Zcode$ into a common isotypic component.
\section{Detailed Code Construction}
\subsection{Two-mode cyclic code}
Construction from the cyclic group $C_4$ recovers the 4-legged cat qubit but also provides a novel two-mode code when choosing a physical representation that is faithful and reducible~\cite{Leghtas2013}. The group $C_n$ has a presentation given by $C_n = \langle x\mid x^n=e\rangle$. The character table of the group $C_4$ appears in \cref{tab:CharacterC4}. We choose as a physical representation $\rho\cong \rho_3\oplus\rho_4$. Here the physical representation remains faithful. The logical representation is chosen to be $\lambda \cong \rho_1\oplus\rho_2$. Using the usual projection formula of \cref{Projection}, one finds the following codewords.
\begin{align}
    \ket{0}_L &\propto \ket{\alpha,\alpha}+\ket{-\alpha,-\alpha}+\ket{i\alpha,-i\alpha}+\ket{-i\alpha,i\alpha}\\
    \ket{1}_L &\propto \ket{\alpha,\alpha}-\ket{-\alpha,-\alpha}+\ket{i\alpha,-i\alpha}-\ket{-i\alpha,i\alpha}
\end{align}
This code is reminiscent of the pair-cat code~\cite{Albert2019} or the concatenated cat-code~\cite{Guillaud2019}. However, unlike the pair-cat code, the error protection does not come from photon number difference between modes. Here the error protection comes from the global phase across the modes between codewords.
\subsection{Bosonic cyclic codes}
Another class of code that we can recover from this construction is the newly published bosonic cyclic code~\cite{Wetherbee2026}. Let's consider the cyclic group $C_{10}$ and its character \cref{tab:CharacterC10}. To define the $\mathbb{Z}_5$ simplex code, we choose as physical representation the reducible faithful representation $\rho\cong\rho_4\oplus\rho_8$ and logical representation $\lambda\cong\rho_1\oplus\rho_2$. Using projection formula of \cref{Projection}, we get the codewords,
\begin{align}
    \ket{0} &= P_{\rho_1}\ket{\alpha,\alpha}\\
    \ket{0} &\propto \sum_{j=0}^4 \ket{\omega^j\alpha,\omega^{2j}\alpha}+\ket{-\omega^j\alpha,-\omega^{2j}\alpha} \\
    \ket{1} &= P_{\rho_2}\ket{\alpha,\alpha}\\
    \ket{1} &\propto \sum_{j=0}^4 \ket{\omega^j\alpha,\omega^{2j}\alpha}-\ket{-\omega^j\alpha,-\omega^{2j}\alpha} \label{Eq:Logical1BosonCyclic}
\end{align}
recovering Eq. (C1) from~\cite{Wetherbee2026}. Knowing that $\{\hat{a}_1,\hat{a}_2\}$ transforms as $\rho^*\cong\rho_8^*\oplus\rho_4^*=\rho_6\oplus\rho_{10}$ and utilizing the fact that the physical representation is reducible, therefore acts independently on each mode, applying a loss operator on a specific mode has the effect of redefining the logical representation. For instance, the operator $\hat{a}_2$ transforms as $\rho_4^*\cong\rho_{10}$. Applying a photon loss on the logical codeword $\ket{1}$ of \cref{Eq:Logical1BosonCyclic} is equivalent, on the representation level to redefining the irreducible representation sector as $\rho_{10}\otimes\rho_2\cong\rho_9$. Under this definition, we recover Eq. (C5) from~\cite{Wetherbee2026}.
\begin{align}
    \ket{0}_{\mathrm{drum}} &\propto P_{\rho_1}\ket{\alpha,\alpha}\\
    \ket{1}_{\mathrm{drum}} &\propto P_{\rho_9}\ket{\alpha,\alpha}
\end{align}
where the logical representation is now $\lambda_{\mathrm{drum}}\cong\rho_1\oplus\rho_9$. A key difference between the two codes are the stabilizer group where we initially had $\mathrm{ker}(\lambda)\cong C_5$ and moved to $\mathrm{ker}(\lambda_{\mathrm{drum}})\cong C_2$.

\section{Character tables}\label{sec:charTables}
\begin{table}[H]
    \centering
    \begin{tabular}{|c|c|c|c|c|c|} \hline
         $Q_8$&$e$&$-e$&$i$&$j$&$k$ \\ \hline
         $\rho_1$&1&1&1&1&1\\ \hline
         $\rho_2$&1&1&-1&1&-1\\ \hline
         $\rho_3$&1&1&1&-1&-1\\ \hline
         $\rho_4$&1&1&-1&-1&1\\ \hline
         $\rho_5$&2&-2&0&0&0\\ \hline
    \end{tabular}
    \caption{Character table for the group $Q_8$}
    \label{tab:CharacterQ8}
\end{table} 
\begin{table}[H]
    \centering
    \begin{tabular}{|c|c|c|c|c|c|c|c|c|} \hline
         $2O$&(e)&$x^2$&$xy$&$y^2$&$x$&$z$&$y$&$xz$  \\ \hline
         $\rho_1$&1&1&1&1&1&1&1&1\\ \hline
         $\rho_2$&1&1&1&1&-1&1&-1&-1 \\ \hline
         $\rho_3$&2&2&-1&2&0&-1&0&0 \\ \hline
         $\rho_4$&2&-2&-1&0&0&1&$\sqrt{2}$&$-\sqrt{2}$ \\ \hline
         $\rho_5$&2&-2&-1&0&0&1&$-\sqrt{2}$&$\sqrt{2}$ \\ \hline
         $\rho_6$&3&3&0&-1&-1&0&1&1 \\ \hline
         $\rho_7$&3&3&0&-1&1&0&-1&-1 \\ \hline
         $\rho_8$&4&-4&1&0&0&-1&0&0 \\ \hline
    \end{tabular}
    \caption{Character table for the group $2O$}
    \label{tab:Character2O}
\end{table}
\begin{table}[H]
    \centering
    \begin{tabular}{|c|c|c|c|c|c|c|c|c|c|c|} \hline
         $C_{10}$&$(e)$&$a^5$&$a^2$&$a^4$&$a^6$&$a^8$&$a$&$a^3$&$a^7$&$a^9$ \\ \hline
         $\rho_1$&1&1&1&1&1&1&1&1&1&1 \\ \hline
         $\rho_2$&1&-1&1&1&1&1&-1&-1&-1&-1\\ \hline
         $\rho_3$&1&1&$\omega^2$&$\omega^3$&$\omega^4$&$\omega$&$\omega^4$&$\omega^3$&$\omega^2$&$\omega$ \\ \hline
         $\rho_4$&1&-1&$\omega^2$&$\omega^3$&$\omega^4$&$\omega$&$-\omega^4$&$-\omega^3$&$-\omega^2$&$-\omega$ \\ \hline
         $\rho_5$&1&1&$\omega^4$&$\omega$&$\omega^3$&$\omega^2$&$\omega^3$&$\omega^4$&$\omega$&$\omega^2$ \\ \hline
         $\rho_6$&1&-1&$\omega^4$&$\omega$&$\omega^3$&$\omega^2$&$-\omega^3$&$-\omega^4$&$-\omega$&$-\omega^2$ \\ \hline
         $\rho_7$&1&1&$\omega$&$\omega^4$&$\omega^2$&$\omega^3$&$\omega^2$&$\omega$&$\omega^4$&$\omega^3$ \\ \hline
         $\rho_8$&1&-1&$\omega$&$\omega^4$&$\omega^2$&$\omega^3$&$-\omega^2$&$-\omega$&$-\omega^4$&$-\omega^3$ \\ \hline
         $\rho_9$&1&1&$\omega^3$&$\omega^2$&$\omega$&$\omega^4$&$\omega$&$\omega^3$&$\omega^2$&$\omega^4$ \\ \hline
         $\rho_{10}$&1&-1&$\omega^3$&$\omega^2$&$\omega$&$\omega^4$&$-\omega$&$-\omega^3$&$-\omega^2$&$-\omega^4$ \\ \hline
    \end{tabular}
    \caption{Character table for the group $C_{10}$}
    \label{tab:CharacterC10}
\end{table}
\begin{table}[H]
    \centering
    \begin{tabular}{|c|c|c|c|c|} \hline
         $C_4$&$(e)$&$a$&$a^2$&$a^3$ \\ \hline
         $\rho_1$&1&1&1&1 \\ \hline
         $\rho_2$&1&i&-1&-i\\ \hline
         $\rho_3$&1&-1&1&-1\\ \hline
         $\rho_4$&1&-i&-1&i \\ \hline
    \end{tabular}
    \caption{Character table for the group $C_4$}
    \label{tab:CharacterC4}
\end{table}
\begin{table}[H]
    \centering
    \begin{tabular}{|c|c|c|c|c|c|c|c|c|} \hline
         $C_8$&$(e)$&$a$&$a^2$&$a^3$&$a^4$&$a^5$&$a^6$&$a^7$ \\ \hline
         $\rho_1$&1&1&1&1&1&1&1&1 \\ \hline
         $\rho_2$&1&$\omega$&$\omega^2$&$\omega^3$&-1&$\omega^5$&$\omega^6$&$\omega^7$\\ \hline
         $\rho_3$&1&$\omega^2$&-1&$\omega^6$&1&$\omega^2$&-1&$\omega^6$ \\ \hline
         $\rho_4$&1&$\omega^3$&$\omega^6$&$\omega$&-1&$\omega^7$&$\omega^2$&$\omega^5$ \\ \hline
         $\rho_5$&1&-1&1&-1&1&-1&1&-1\\ \hline
         $\rho_6$&1&$\omega^5$&$\omega^2$&$\omega^7$&-1&$\omega$&$\omega^6$&$\omega^3$ \\ \hline
         $\rho_7$&1&$\omega^6$&-1&$\omega^2$&1&$\omega^6$&-1&$\omega^2$ \\ \hline
         $\rho_8$&1&$\omega^7$&$\omega^6$&$\omega^5$&-1&$\omega^3$&$\omega^2$&$\omega$ \\ \hline
    \end{tabular}
    \caption{Character table for the group $C_8$}
    \label{tab:CharacterC8}
\end{table}
\section{Codewords in the same isotypic component}\label{app:ModEncoding}
As stated in the main text, the encoding operator can be written as
\begin{equation}
    V_G = P_{\rho_m}\frac{\ket{\balpha}\bra{0}}{\sqrt{\bra{\balpha}P_{\rho_m} \ket{\balpha}}}  + P_{\rho_n}\frac{\ket{\balpha}\bra{1}}{\sqrt{\bra{\balpha}P_{\rho_n} \ket{\balpha}}}  .
\end{equation}
However, when $m=n$ both codewords become equal and one needs to slightly modify the encoding. Instead of automatically achieving orthogonality through the irreducible representation, we now need to fix it through a choice of initial state. We need two initial states $\ket{\balpha_0},\ket{\balpha_1}$ such that their projection onto isotypic sectors are linearly independent. One can then find an orthonormal basis for the code space $\mathcal C = \mathrm{span}\{P_{\rho_m}\ket{\balpha_0},P_{\rho_m}\ket{\balpha_1}\}$ through, for example, a Gram-Schmidt procedure. 

\section{Near-optimal fidelity optimization}\label{app:NearOptFidOptimization}
In order to confirm and strengthen the statement made in \cref{sec:InitState}, we reproduce here panel b) of \cref{fig:Q8-24,fig:2O-12,fig:2O-3} side-by-side with the explicit calculation of \cref{Eq:TammesQuantum} as a function of the energy ratio between the first mode and the total energy and the phase difference between the mode for the optimized energy found in panel c) of \cref{fig:Q8-24,fig:2O-12,fig:2O-3}. This serves as a qualitative comparison between the general shape of the two metrics. We focus mainly on the general appearance of the two plots and the apparent difference between the two states that optimize both metric. 
\begin{figure*}[tp]
    \centering
    \includegraphics[width=0.85\textwidth]{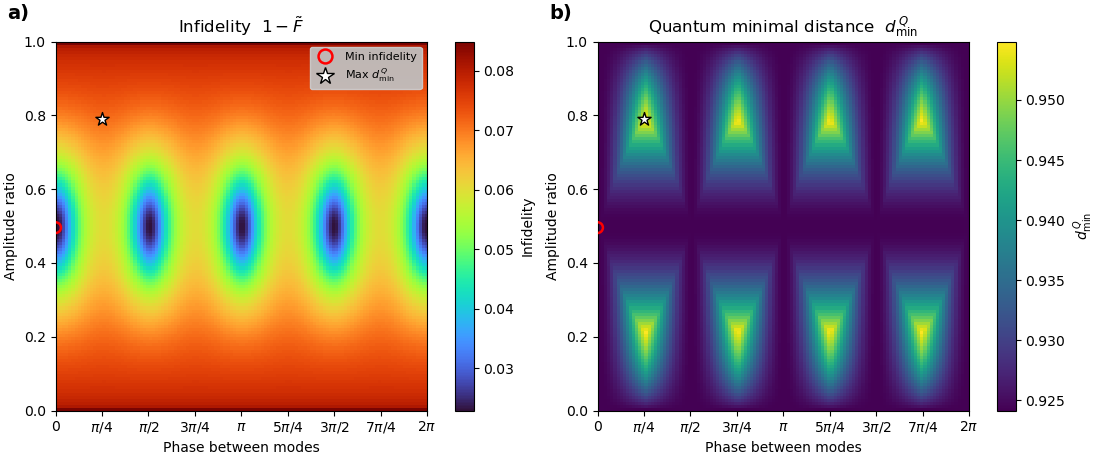}
    \caption{Comparison of the amplitude-damping infidelity and the quantum minimal
  distance over the seed-state parameters at fixed energy $|\alpha|^2 = 0.528$ for the group $Q_8$ and logical representation $\lambda\cong\rho_2\oplus\rho_4$.}   
\end{figure*}
\begin{figure*}[tp]
    \centering
    \includegraphics[width=0.85\textwidth]{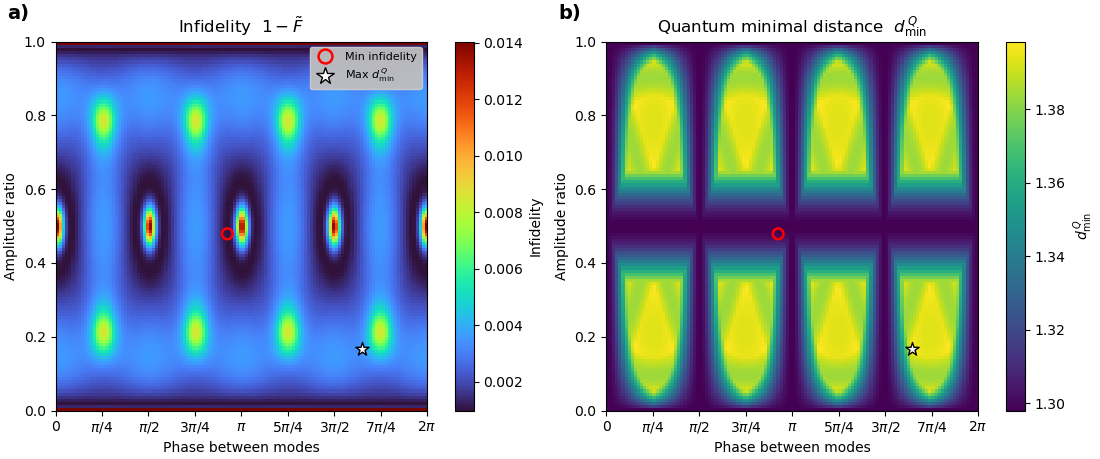}
    \caption{Comparison of the amplitude-damping infidelity and the quantum minimal
  distance over the seed-state parameters at fixed energy $|\alpha|^2 = 6.31$ for the group $2O$ and logical representation $\lambda\cong\rho_1\oplus\rho_2$.}   
\end{figure*}
\begin{figure*}[tp]
    \centering
    \includegraphics[width=0.85\textwidth]{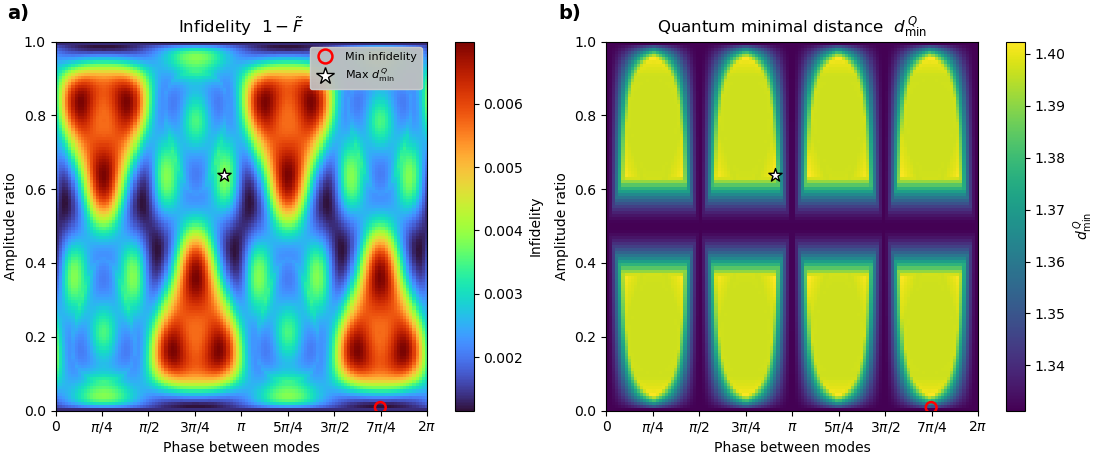}
    \caption{Comparison of the amplitude-damping infidelity and the quantum minimal
  distance over the seed-state parameters at fixed energy $|\alpha|^2 = 7.42$ for the group $2O$ and logical representation $\lambda\cong\rho_3$.}   
\end{figure*}
\nocite{*}
\clearpage

\begin{thebibliography}{47}%
\makeatletter
\providecommand \@ifxundefined [1]{%
 \@ifx{#1\undefined}
}%
\providecommand \@ifnum [1]{%
 \ifnum #1\expandafter \@firstoftwo
 \else \expandafter \@secondoftwo
 \fi
}%
\providecommand \@ifx [1]{%
 \ifx #1\expandafter \@firstoftwo
 \else \expandafter \@secondoftwo
 \fi
}%
\providecommand \natexlab [1]{#1}%
\providecommand \enquote  [1]{``#1''}%
\providecommand \bibnamefont  [1]{#1}%
\providecommand \bibfnamefont [1]{#1}%
\providecommand \citenamefont [1]{#1}%
\providecommand \href@noop [0]{\@secondoftwo}%
\providecommand \href [0]{\begingroup \@sanitize@url \@href}%
\providecommand \@href[1]{\@@startlink{#1}\@@href}%
\providecommand \@@href[1]{\endgroup#1\@@endlink}%
\providecommand \@sanitize@url [0]{\catcode `\\12\catcode `\$12\catcode `\&12\catcode `\#12\catcode `\^12\catcode `\_12\catcode `\%12\relax}%
\providecommand \@@startlink[1]{}%
\providecommand \@@endlink[0]{}%
\providecommand \url  [0]{\begingroup\@sanitize@url \@url }%
\providecommand \@url [1]{\endgroup\@href {#1}{\urlprefix }}%
\providecommand \urlprefix  [0]{URL }%
\providecommand \Eprint [0]{\href }%
\providecommand \doibase [0]{https://doi.org/}%
\providecommand \selectlanguage [0]{\@gobble}%
\providecommand \bibinfo  [0]{\@secondoftwo}%
\providecommand \bibfield  [0]{\@secondoftwo}%
\providecommand \translation [1]{[#1]}%
\providecommand \BibitemOpen [0]{}%
\providecommand \bibitemStop [0]{}%
\providecommand \bibitemNoStop [0]{.\EOS\space}%
\providecommand \EOS [0]{\spacefactor3000\relax}%
\providecommand \BibitemShut  [1]{\csname bibitem#1\endcsname}%
\let\auto@bib@innerbib\@empty
\bibitem [{\citenamefont {Corcoles}\ \emph {et~al.}(2020)\citenamefont {Corcoles}, \citenamefont {Kandala}, \citenamefont {Javadi-Abhari}, \citenamefont {McClure}, \citenamefont {Cross}, \citenamefont {Temme}, \citenamefont {Nation}, \citenamefont {Steffen},\ and\ \citenamefont {Gambetta}}]{Corcoles2020}%
  \BibitemOpen
  \bibfield  {author} {\bibinfo {author} {\bibfnamefont {A.~D.}\ \bibnamefont {Corcoles}}, \bibinfo {author} {\bibfnamefont {A.}~\bibnamefont {Kandala}}, \bibinfo {author} {\bibfnamefont {A.}~\bibnamefont {Javadi-Abhari}}, \bibinfo {author} {\bibfnamefont {D.~T.}\ \bibnamefont {McClure}}, \bibinfo {author} {\bibfnamefont {A.~W.}\ \bibnamefont {Cross}}, \bibinfo {author} {\bibfnamefont {K.}~\bibnamefont {Temme}}, \bibinfo {author} {\bibfnamefont {P.~D.}\ \bibnamefont {Nation}}, \bibinfo {author} {\bibfnamefont {M.}~\bibnamefont {Steffen}},\ and\ \bibinfo {author} {\bibfnamefont {J.~M.}\ \bibnamefont {Gambetta}},\ }\bibfield  {title} {\bibinfo {title} {Challenges and opportunities of near-term quantum computing systems},\ }\href {https://doi.org/10.1109/jproc.2019.2954005} {\bibfield  {journal} {\bibinfo  {journal} {Proceedings of the IEEE}\ }\textbf {\bibinfo {volume} {108}},\ \bibinfo {pages} {1338} (\bibinfo {year} {2020})}\BibitemShut {NoStop}%
\bibitem [{\citenamefont {Fowler}\ \emph {et~al.}(2012)\citenamefont {Fowler}, \citenamefont {Mariantoni}, \citenamefont {Martinis},\ and\ \citenamefont {Cleland}}]{Fowler2012}%
  \BibitemOpen
  \bibfield  {author} {\bibinfo {author} {\bibfnamefont {A.~G.}\ \bibnamefont {Fowler}}, \bibinfo {author} {\bibfnamefont {M.}~\bibnamefont {Mariantoni}}, \bibinfo {author} {\bibfnamefont {J.~M.}\ \bibnamefont {Martinis}},\ and\ \bibinfo {author} {\bibfnamefont {A.~N.}\ \bibnamefont {Cleland}},\ }\bibfield  {title} {\bibinfo {title} {Surface codes: Towards practical large-scale quantum computation},\ }\href {https://doi.org/10.1103/physreva.86.032324} {\bibfield  {journal} {\bibinfo  {journal} {Physical Review A}\ }\textbf {\bibinfo {volume} {86}},\ \bibinfo {pages} {032324} (\bibinfo {year} {2012})}\BibitemShut {NoStop}%
\bibitem [{\citenamefont {Cai}\ \emph {et~al.}(2021)\citenamefont {Cai}, \citenamefont {Ma}, \citenamefont {Wang}, \citenamefont {Zou},\ and\ \citenamefont {Sun}}]{Cai2021}%
  \BibitemOpen
  \bibfield  {author} {\bibinfo {author} {\bibfnamefont {W.}~\bibnamefont {Cai}}, \bibinfo {author} {\bibfnamefont {Y.}~\bibnamefont {Ma}}, \bibinfo {author} {\bibfnamefont {W.}~\bibnamefont {Wang}}, \bibinfo {author} {\bibfnamefont {C.-L.}\ \bibnamefont {Zou}},\ and\ \bibinfo {author} {\bibfnamefont {L.}~\bibnamefont {Sun}},\ }\bibfield  {title} {\bibinfo {title} {Bosonic quantum error correction codes in superconducting quantum circuits},\ }\href {https://doi.org/10.1016/j.fmre.2020.12.006} {\bibfield  {journal} {\bibinfo  {journal} {Fundamental Research}\ }\textbf {\bibinfo {volume} {1}},\ \bibinfo {pages} {50} (\bibinfo {year} {2021})}\BibitemShut {NoStop}%
\bibitem [{\citenamefont {Gottesman}\ \emph {et~al.}(2001)\citenamefont {Gottesman}, \citenamefont {Kitaev},\ and\ \citenamefont {Preskill}}]{Gottesman2001}%
  \BibitemOpen
  \bibfield  {author} {\bibinfo {author} {\bibfnamefont {D.}~\bibnamefont {Gottesman}}, \bibinfo {author} {\bibfnamefont {A.}~\bibnamefont {Kitaev}},\ and\ \bibinfo {author} {\bibfnamefont {J.}~\bibnamefont {Preskill}},\ }\bibfield  {title} {\bibinfo {title} {Encoding a qubit in an oscillator},\ }\href {https://doi.org/10.1103/physreva.64.012310} {\bibfield  {journal} {\bibinfo  {journal} {Physical Review A}\ }\textbf {\bibinfo {volume} {64}},\ \bibinfo {pages} {012310} (\bibinfo {year} {2001})}\BibitemShut {NoStop}%
\bibitem [{\citenamefont {Leghtas}\ \emph {et~al.}(2013)\citenamefont {Leghtas}, \citenamefont {Kirchmair}, \citenamefont {Vlastakis}, \citenamefont {Schoelkopf}, \citenamefont {Devoret},\ and\ \citenamefont {Mirrahimi}}]{Leghtas2013}%
  \BibitemOpen
  \bibfield  {author} {\bibinfo {author} {\bibfnamefont {Z.}~\bibnamefont {Leghtas}}, \bibinfo {author} {\bibfnamefont {G.}~\bibnamefont {Kirchmair}}, \bibinfo {author} {\bibfnamefont {B.}~\bibnamefont {Vlastakis}}, \bibinfo {author} {\bibfnamefont {R.~J.}\ \bibnamefont {Schoelkopf}}, \bibinfo {author} {\bibfnamefont {M.~H.}\ \bibnamefont {Devoret}},\ and\ \bibinfo {author} {\bibfnamefont {M.}~\bibnamefont {Mirrahimi}},\ }\bibfield  {title} {\bibinfo {title} {Hardware-efficient autonomous quantum memory protection},\ }\href {https://doi.org/10.1103/physrevlett.111.120501} {\bibfield  {journal} {\bibinfo  {journal} {Physical Review Letters}\ }\textbf {\bibinfo {volume} {111}},\ \bibinfo {pages} {120501} (\bibinfo {year} {2013})}\BibitemShut {NoStop}%
\bibitem [{\citenamefont {Grimsmo}\ \emph {et~al.}(2020)\citenamefont {Grimsmo}, \citenamefont {Combes},\ and\ \citenamefont {Baragiola}}]{Grimsmo2020}%
  \BibitemOpen
  \bibfield  {author} {\bibinfo {author} {\bibfnamefont {A.~L.}\ \bibnamefont {Grimsmo}}, \bibinfo {author} {\bibfnamefont {J.}~\bibnamefont {Combes}},\ and\ \bibinfo {author} {\bibfnamefont {B.~Q.}\ \bibnamefont {Baragiola}},\ }\bibfield  {title} {\bibinfo {title} {Quantum computing with rotation-symmetric bosonic codes},\ }\href {https://doi.org/10.1103/physrevx.10.011058} {\bibfield  {journal} {\bibinfo  {journal} {Physical Review X}\ }\textbf {\bibinfo {volume} {10}},\ \bibinfo {pages} {011058} (\bibinfo {year} {2020})}\BibitemShut {NoStop}%
\bibitem [{\citenamefont {Jain}\ \emph {et~al.}(2024)\citenamefont {Jain}, \citenamefont {Iosue}, \citenamefont {Barg},\ and\ \citenamefont {Albert}}]{Jain2024}%
  \BibitemOpen
  \bibfield  {author} {\bibinfo {author} {\bibfnamefont {S.~P.}\ \bibnamefont {Jain}}, \bibinfo {author} {\bibfnamefont {J.~T.}\ \bibnamefont {Iosue}}, \bibinfo {author} {\bibfnamefont {A.}~\bibnamefont {Barg}},\ and\ \bibinfo {author} {\bibfnamefont {V.~V.}\ \bibnamefont {Albert}},\ }\bibfield  {title} {\bibinfo {title} {Quantum spherical codes},\ }\href {https://doi.org/10.1038/s41567-024-02496-y} {\bibfield  {journal} {\bibinfo  {journal} {Nature Physics}\ }\textbf {\bibinfo {volume} {20}},\ \bibinfo {pages} {1300} (\bibinfo {year} {2024})}\BibitemShut {NoStop}%
\bibitem [{\citenamefont {Gross}(2021)}]{Gross2021}%
  \BibitemOpen
  \bibfield  {author} {\bibinfo {author} {\bibfnamefont {J.~A.}\ \bibnamefont {Gross}},\ }\bibfield  {title} {\bibinfo {title} {Designing codes around interactions: The case of a spin},\ }\href {https://doi.org/10.1103/PhysRevLett.127.010504} {\bibfield  {journal} {\bibinfo  {journal} {Phys. Rev. Lett.}\ }\textbf {\bibinfo {volume} {127}},\ \bibinfo {pages} {010504} (\bibinfo {year} {2021})}\BibitemShut {NoStop}%
\bibitem [{\citenamefont {Denys}\ and\ \citenamefont {Leverrier}(2023)}]{Denys2023}%
  \BibitemOpen
  \bibfield  {author} {\bibinfo {author} {\bibfnamefont {A.}~\bibnamefont {Denys}}\ and\ \bibinfo {author} {\bibfnamefont {A.}~\bibnamefont {Leverrier}},\ }\bibfield  {title} {\bibinfo {title} {The 2t-qutrit, a two-mode bosonic qutrit},\ }\href {https://doi.org/10.22331/q-2023-06-05-1032} {\bibfield  {journal} {\bibinfo  {journal} {Quantum}\ }\textbf {\bibinfo {volume} {7}},\ \bibinfo {pages} {1032} (\bibinfo {year} {2023})}\BibitemShut {NoStop}%
\bibitem [{\citenamefont {Denys}\ and\ \citenamefont {Leverrier}(2024)}]{Covariant2024}%
  \BibitemOpen
  \bibfield  {author} {\bibinfo {author} {\bibfnamefont {A.}~\bibnamefont {Denys}}\ and\ \bibinfo {author} {\bibfnamefont {A.}~\bibnamefont {Leverrier}},\ }\bibfield  {title} {\bibinfo {title} {Quantum error-correcting codes with a covariant encoding},\ }\href {https://doi.org/10.1103/PhysRevLett.133.240603} {\bibfield  {journal} {\bibinfo  {journal} {Phys. Rev. Lett.}\ }\textbf {\bibinfo {volume} {133}},\ \bibinfo {pages} {240603} (\bibinfo {year} {2024})}\BibitemShut {NoStop}%
\bibitem [{\citenamefont {Leverrier}(2026)}]{Leverrier2026}%
  \BibitemOpen
  \bibfield  {author} {\bibinfo {author} {\bibfnamefont {A.}~\bibnamefont {Leverrier}},\ }\bibfield  {title} {\bibinfo {title} {Bosonic quantum fourier codes},\ }\href {https://doi.org/10.22331/q-2026-02-09-2000} {\bibfield  {journal} {\bibinfo  {journal} {Quantum}\ }\textbf {\bibinfo {volume} {10}},\ \bibinfo {pages} {2000} (\bibinfo {year} {2026})}\BibitemShut {NoStop}%
\bibitem [{\citenamefont {Ahmed}\ \emph {et~al.}(2026)\citenamefont {Ahmed}, \citenamefont {Udupa},\ and\ \citenamefont {Ferrini}}]{Ahmed2026}%
  \BibitemOpen
  \bibfield  {author} {\bibinfo {author} {\bibfnamefont {R.~G.}\ \bibnamefont {Ahmed}}, \bibinfo {author} {\bibfnamefont {A.}~\bibnamefont {Udupa}},\ and\ \bibinfo {author} {\bibfnamefont {G.}~\bibnamefont {Ferrini}},\ }\bibfield  {title} {\bibinfo {title} {Multimode rotationally symmetric bosonic codes from group-theoretic construction},\ }\href {https://doi.org/10.1103/xt48-fxt5} {\bibfield  {journal} {\bibinfo  {journal} {Phys. Rev. Res.}\ }\textbf {\bibinfo {volume} {8}},\ \bibinfo {pages} {023194} (\bibinfo {year} {2026})}\BibitemShut {NoStop}%
\bibitem [{\citenamefont {Kubischta}\ and\ \citenamefont {Teixeira}(2025)}]{Kubischta2025}%
  \BibitemOpen
  \bibfield  {author} {\bibinfo {author} {\bibfnamefont {E.}~\bibnamefont {Kubischta}}\ and\ \bibinfo {author} {\bibfnamefont {I.}~\bibnamefont {Teixeira}},\ }\bibfield  {title} {\bibinfo {title} {Intrinsic quantum codes},\ }\bibfield  {journal} {\bibinfo  {journal} {arXiv}\ }\href {https://doi.org/10.48550/ARXIV.2511.14840} {10.48550/ARXIV.2511.14840} (\bibinfo {year} {2025}),\ \Eprint {https://arxiv.org/abs/2511.14840} {arXiv:2511.14840 [quant-ph]} \BibitemShut {NoStop}%
\bibitem [{\citenamefont {Ofek}\ \emph {et~al.}(2016)\citenamefont {Ofek}, \citenamefont {Petrenko}, \citenamefont {Heeres}, \citenamefont {Reinhold}, \citenamefont {Leghtas}, \citenamefont {Vlastakis}, \citenamefont {Liu}, \citenamefont {Frunzio}, \citenamefont {Girvin}, \citenamefont {Jiang}, \citenamefont {Mirrahimi}, \citenamefont {Devoret},\ and\ \citenamefont {Schoelkopf}}]{Ofek2016}%
  \BibitemOpen
  \bibfield  {author} {\bibinfo {author} {\bibfnamefont {N.}~\bibnamefont {Ofek}}, \bibinfo {author} {\bibfnamefont {A.}~\bibnamefont {Petrenko}}, \bibinfo {author} {\bibfnamefont {R.}~\bibnamefont {Heeres}}, \bibinfo {author} {\bibfnamefont {P.}~\bibnamefont {Reinhold}}, \bibinfo {author} {\bibfnamefont {Z.}~\bibnamefont {Leghtas}}, \bibinfo {author} {\bibfnamefont {B.}~\bibnamefont {Vlastakis}}, \bibinfo {author} {\bibfnamefont {Y.}~\bibnamefont {Liu}}, \bibinfo {author} {\bibfnamefont {L.}~\bibnamefont {Frunzio}}, \bibinfo {author} {\bibfnamefont {S.~M.}\ \bibnamefont {Girvin}}, \bibinfo {author} {\bibfnamefont {L.}~\bibnamefont {Jiang}}, \bibinfo {author} {\bibfnamefont {M.}~\bibnamefont {Mirrahimi}}, \bibinfo {author} {\bibfnamefont {M.~H.}\ \bibnamefont {Devoret}},\ and\ \bibinfo {author} {\bibfnamefont {R.~J.}\ \bibnamefont {Schoelkopf}},\ }\bibfield  {title} {\bibinfo {title} {Extending the lifetime of a quantum bit with error correction in superconducting circuits},\ }\href
  {https://doi.org/10.1038/nature18949} {\bibfield  {journal} {\bibinfo  {journal} {Nature}\ }\textbf {\bibinfo {volume} {536}},\ \bibinfo {pages} {441} (\bibinfo {year} {2016})}\BibitemShut {NoStop}%
\bibitem [{\citenamefont {Sivak}\ \emph {et~al.}(2023)\citenamefont {Sivak}, \citenamefont {Eickbusch}, \citenamefont {Royer}, \citenamefont {Singh}, \citenamefont {Tsioutsios}, \citenamefont {Ganjam}, \citenamefont {Miano}, \citenamefont {Brock}, \citenamefont {Ding}, \citenamefont {Frunzio}, \citenamefont {Girvin}, \citenamefont {Schoelkopf},\ and\ \citenamefont {Devoret}}]{Sivak2023}%
  \BibitemOpen
  \bibfield  {author} {\bibinfo {author} {\bibfnamefont {V.~V.}\ \bibnamefont {Sivak}}, \bibinfo {author} {\bibfnamefont {A.}~\bibnamefont {Eickbusch}}, \bibinfo {author} {\bibfnamefont {B.}~\bibnamefont {Royer}}, \bibinfo {author} {\bibfnamefont {S.}~\bibnamefont {Singh}}, \bibinfo {author} {\bibfnamefont {I.}~\bibnamefont {Tsioutsios}}, \bibinfo {author} {\bibfnamefont {S.}~\bibnamefont {Ganjam}}, \bibinfo {author} {\bibfnamefont {A.}~\bibnamefont {Miano}}, \bibinfo {author} {\bibfnamefont {B.~L.}\ \bibnamefont {Brock}}, \bibinfo {author} {\bibfnamefont {A.~Z.}\ \bibnamefont {Ding}}, \bibinfo {author} {\bibfnamefont {L.}~\bibnamefont {Frunzio}}, \bibinfo {author} {\bibfnamefont {S.~M.}\ \bibnamefont {Girvin}}, \bibinfo {author} {\bibfnamefont {R.~J.}\ \bibnamefont {Schoelkopf}},\ and\ \bibinfo {author} {\bibfnamefont {M.~H.}\ \bibnamefont {Devoret}},\ }\bibfield  {title} {\bibinfo {title} {Real-time quantum error correction beyond break-even},\ }\href {https://doi.org/10.1038/s41586-023-05782-6} {\bibfield
  {journal} {\bibinfo  {journal} {Nature}\ }\textbf {\bibinfo {volume} {616}},\ \bibinfo {pages} {50} (\bibinfo {year} {2023})}\BibitemShut {NoStop}%
\bibitem [{\citenamefont {Ni}\ \emph {et~al.}(2023)\citenamefont {Ni}, \citenamefont {Li}, \citenamefont {Deng}, \citenamefont {Cai}, \citenamefont {Zhang}, \citenamefont {Wang}, \citenamefont {Yang}, \citenamefont {Yu}, \citenamefont {Yan}, \citenamefont {Liu}, \citenamefont {Zou}, \citenamefont {Sun}, \citenamefont {Zheng}, \citenamefont {Xu},\ and\ \citenamefont {Yu}}]{Ni2023}%
  \BibitemOpen
  \bibfield  {author} {\bibinfo {author} {\bibfnamefont {Z.}~\bibnamefont {Ni}}, \bibinfo {author} {\bibfnamefont {S.}~\bibnamefont {Li}}, \bibinfo {author} {\bibfnamefont {X.}~\bibnamefont {Deng}}, \bibinfo {author} {\bibfnamefont {Y.}~\bibnamefont {Cai}}, \bibinfo {author} {\bibfnamefont {L.}~\bibnamefont {Zhang}}, \bibinfo {author} {\bibfnamefont {W.}~\bibnamefont {Wang}}, \bibinfo {author} {\bibfnamefont {Z.-B.}\ \bibnamefont {Yang}}, \bibinfo {author} {\bibfnamefont {H.}~\bibnamefont {Yu}}, \bibinfo {author} {\bibfnamefont {F.}~\bibnamefont {Yan}}, \bibinfo {author} {\bibfnamefont {S.}~\bibnamefont {Liu}}, \bibinfo {author} {\bibfnamefont {C.-L.}\ \bibnamefont {Zou}}, \bibinfo {author} {\bibfnamefont {L.}~\bibnamefont {Sun}}, \bibinfo {author} {\bibfnamefont {S.-B.}\ \bibnamefont {Zheng}}, \bibinfo {author} {\bibfnamefont {Y.}~\bibnamefont {Xu}},\ and\ \bibinfo {author} {\bibfnamefont {D.}~\bibnamefont {Yu}},\ }\bibfield  {title} {\bibinfo {title} {Beating the break-even point with a
  discrete-variable-encoded logical qubit},\ }\href {https://doi.org/10.1038/s41586-023-05784-4} {\bibfield  {journal} {\bibinfo  {journal} {Nature}\ }\textbf {\bibinfo {volume} {616}},\ \bibinfo {pages} {56} (\bibinfo {year} {2023})}\BibitemShut {NoStop}%
\bibitem [{\citenamefont {Brock}\ \emph {et~al.}(2025)\citenamefont {Brock}, \citenamefont {Singh}, \citenamefont {Eickbusch}, \citenamefont {Sivak}, \citenamefont {Ding}, \citenamefont {Frunzio}, \citenamefont {Girvin},\ and\ \citenamefont {Devoret}}]{Brock2025}%
  \BibitemOpen
  \bibfield  {author} {\bibinfo {author} {\bibfnamefont {B.~L.}\ \bibnamefont {Brock}}, \bibinfo {author} {\bibfnamefont {S.}~\bibnamefont {Singh}}, \bibinfo {author} {\bibfnamefont {A.}~\bibnamefont {Eickbusch}}, \bibinfo {author} {\bibfnamefont {V.~V.}\ \bibnamefont {Sivak}}, \bibinfo {author} {\bibfnamefont {A.~Z.}\ \bibnamefont {Ding}}, \bibinfo {author} {\bibfnamefont {L.}~\bibnamefont {Frunzio}}, \bibinfo {author} {\bibfnamefont {S.~M.}\ \bibnamefont {Girvin}},\ and\ \bibinfo {author} {\bibfnamefont {M.~H.}\ \bibnamefont {Devoret}},\ }\bibfield  {title} {\bibinfo {title} {Quantum error correction of qudits beyond break-even},\ }\href {https://doi.org/10.1038/s41586-025-08899-y} {\bibfield  {journal} {\bibinfo  {journal} {Nature}\ }\textbf {\bibinfo {volume} {641}},\ \bibinfo {pages} {612} (\bibinfo {year} {2025})}\BibitemShut {NoStop}%
\bibitem [{\citenamefont {Fulton}\ and\ \citenamefont {Harris}(2004)}]{Fulton2004}%
  \BibitemOpen
  \bibfield  {author} {\bibinfo {author} {\bibfnamefont {W.}~\bibnamefont {Fulton}}\ and\ \bibinfo {author} {\bibfnamefont {J.}~\bibnamefont {Harris}},\ }\href@noop {} {\emph {\bibinfo {title} {Representation Theory}}},\ Graduate Texts in Mathematics\ (\bibinfo  {publisher} {Springer New York},\ \bibinfo {year} {2004})\ p.\ \bibinfo {pages} {551}\BibitemShut {NoStop}%
\bibitem [{\citenamefont {Hayden}\ \emph {et~al.}(2017)\citenamefont {Hayden}, \citenamefont {Nezami}, \citenamefont {Popescu},\ and\ \citenamefont {Salton}}]{Hayden2017}%
  \BibitemOpen
  \bibfield  {author} {\bibinfo {author} {\bibfnamefont {P.}~\bibnamefont {Hayden}}, \bibinfo {author} {\bibfnamefont {S.}~\bibnamefont {Nezami}}, \bibinfo {author} {\bibfnamefont {S.}~\bibnamefont {Popescu}},\ and\ \bibinfo {author} {\bibfnamefont {G.}~\bibnamefont {Salton}},\ }\bibfield  {title} {\bibinfo {title} {Error correction of quantum reference frame information},\ }\bibfield  {journal} {\bibinfo  {journal} {PRX Quantum}\ }\href {https://doi.org/10.1103/PRXQUANTUM.2.010326} {10.1103/PRXQUANTUM.2.010326} (\bibinfo {year} {2017})\BibitemShut {NoStop}%
\bibitem [{\citenamefont {Faist}\ \emph {et~al.}(2020)\citenamefont {Faist}, \citenamefont {Nezami}, \citenamefont {Albert}, \citenamefont {Salton}, \citenamefont {Pastawski}, \citenamefont {Hayden},\ and\ \citenamefont {Preskill}}]{Faist2020}%
  \BibitemOpen
  \bibfield  {author} {\bibinfo {author} {\bibfnamefont {P.}~\bibnamefont {Faist}}, \bibinfo {author} {\bibfnamefont {S.}~\bibnamefont {Nezami}}, \bibinfo {author} {\bibfnamefont {V.~V.}\ \bibnamefont {Albert}}, \bibinfo {author} {\bibfnamefont {G.}~\bibnamefont {Salton}}, \bibinfo {author} {\bibfnamefont {F.}~\bibnamefont {Pastawski}}, \bibinfo {author} {\bibfnamefont {P.}~\bibnamefont {Hayden}},\ and\ \bibinfo {author} {\bibfnamefont {J.}~\bibnamefont {Preskill}},\ }\bibfield  {title} {\bibinfo {title} {Continuous symmetries and approximate quantum error correction},\ }\href {https://doi.org/10.1103/physrevx.10.041018} {\bibfield  {journal} {\bibinfo  {journal} {Physical Review X}\ }\textbf {\bibinfo {volume} {10}},\ \bibinfo {pages} {041018} (\bibinfo {year} {2020})}\BibitemShut {NoStop}%
\bibitem [{\citenamefont {Liu}\ and\ \citenamefont {Zhou}(2021)}]{Liu2021}%
  \BibitemOpen
  \bibfield  {author} {\bibinfo {author} {\bibfnamefont {Z.-W.}\ \bibnamefont {Liu}}\ and\ \bibinfo {author} {\bibfnamefont {S.}~\bibnamefont {Zhou}},\ }\bibfield  {title} {\bibinfo {title} {Quantum error correction meets continuous symmetries: fundamental trade-offs and case studies},\ }\href {https://arxiv.org/abs/2111.06360} {\bibfield  {journal} {\bibinfo  {journal} {arXiv e-prints}\ } (\bibinfo {year} {2021})}\BibitemShut {NoStop}%
\bibitem [{\citenamefont {Kong}\ and\ \citenamefont {Liu}(2021)}]{Kong2021}%
  \BibitemOpen
  \bibfield  {author} {\bibinfo {author} {\bibfnamefont {L.}~\bibnamefont {Kong}}\ and\ \bibinfo {author} {\bibfnamefont {Z.-W.}\ \bibnamefont {Liu}},\ }\bibfield  {title} {\bibinfo {title} {Near-optimal covariant quantum error-correcting codes from random unitaries with symmetries},\ }\bibfield  {journal} {\bibinfo  {journal} {PRX Quantum}\ }\href {https://doi.org/10.1103/PRXQuantum.3.020314} {10.1103/PRXQuantum.3.020314} (\bibinfo {year} {2021})\BibitemShut {NoStop}%
\bibitem [{\citenamefont {Zhou}\ \emph {et~al.}(2021)\citenamefont {Zhou}, \citenamefont {Liu},\ and\ \citenamefont {Jiang}}]{Zhou2021}%
  \BibitemOpen
  \bibfield  {author} {\bibinfo {author} {\bibfnamefont {S.}~\bibnamefont {Zhou}}, \bibinfo {author} {\bibfnamefont {Z.-W.}\ \bibnamefont {Liu}},\ and\ \bibinfo {author} {\bibfnamefont {L.}~\bibnamefont {Jiang}},\ }\bibfield  {title} {\bibinfo {title} {New perspectives on covariant quantum error correction},\ }\href {https://doi.org/10.22331/q-2021-08-09-521} {\bibfield  {journal} {\bibinfo  {journal} {Quantum}\ }\textbf {\bibinfo {volume} {5}},\ \bibinfo {pages} {521} (\bibinfo {year} {2021})}\BibitemShut {NoStop}%
\bibitem [{\citenamefont {Knill}\ \emph {et~al.}(2000)\citenamefont {Knill}, \citenamefont {Laflamme},\ and\ \citenamefont {Viola}}]{Knill2000}%
  \BibitemOpen
  \bibfield  {author} {\bibinfo {author} {\bibfnamefont {E.}~\bibnamefont {Knill}}, \bibinfo {author} {\bibfnamefont {R.}~\bibnamefont {Laflamme}},\ and\ \bibinfo {author} {\bibfnamefont {L.}~\bibnamefont {Viola}},\ }\bibfield  {title} {\bibinfo {title} {Theory of quantum error correction for general noise},\ }\href {https://doi.org/10.1103/physrevlett.84.2525} {\bibfield  {journal} {\bibinfo  {journal} {Physical Review Letters}\ }\textbf {\bibinfo {volume} {84}},\ \bibinfo {pages} {2525} (\bibinfo {year} {2000})}\BibitemShut {NoStop}%
\bibitem [{\citenamefont {Knill}\ and\ \citenamefont {Laflamme}(1997)}]{Knill1997}%
  \BibitemOpen
  \bibfield  {author} {\bibinfo {author} {\bibfnamefont {E.}~\bibnamefont {Knill}}\ and\ \bibinfo {author} {\bibfnamefont {R.}~\bibnamefont {Laflamme}},\ }\bibfield  {title} {\bibinfo {title} {Theory of quantum error-correcting codes},\ }\href {https://doi.org/10.1103/physreva.55.900} {\bibfield  {journal} {\bibinfo  {journal} {Physical Review A}\ }\textbf {\bibinfo {volume} {55}},\ \bibinfo {pages} {900} (\bibinfo {year} {1997})}\BibitemShut {NoStop}%
\bibitem [{\citenamefont {Albert}\ \emph {et~al.}(2018)\citenamefont {Albert}, \citenamefont {Noh}, \citenamefont {Duivenvoorden}, \citenamefont {Young}, \citenamefont {Brierley}, \citenamefont {Reinhold}, \citenamefont {Vuillot}, \citenamefont {Li}, \citenamefont {Shen}, \citenamefont {Girvin}, \citenamefont {Terhal},\ and\ \citenamefont {Jiang}}]{Albert2018}%
  \BibitemOpen
  \bibfield  {author} {\bibinfo {author} {\bibfnamefont {V.~V.}\ \bibnamefont {Albert}}, \bibinfo {author} {\bibfnamefont {K.}~\bibnamefont {Noh}}, \bibinfo {author} {\bibfnamefont {K.}~\bibnamefont {Duivenvoorden}}, \bibinfo {author} {\bibfnamefont {D.~J.}\ \bibnamefont {Young}}, \bibinfo {author} {\bibfnamefont {R.~T.}\ \bibnamefont {Brierley}}, \bibinfo {author} {\bibfnamefont {P.}~\bibnamefont {Reinhold}}, \bibinfo {author} {\bibfnamefont {C.}~\bibnamefont {Vuillot}}, \bibinfo {author} {\bibfnamefont {L.}~\bibnamefont {Li}}, \bibinfo {author} {\bibfnamefont {C.}~\bibnamefont {Shen}}, \bibinfo {author} {\bibfnamefont {S.~M.}\ \bibnamefont {Girvin}}, \bibinfo {author} {\bibfnamefont {B.~M.}\ \bibnamefont {Terhal}},\ and\ \bibinfo {author} {\bibfnamefont {L.}~\bibnamefont {Jiang}},\ }\bibfield  {title} {\bibinfo {title} {Performance and structure of single-mode bosonic codes},\ }\href {https://doi.org/10.1103/physreva.97.032346} {\bibfield  {journal} {\bibinfo  {journal} {Physical Review A}\ }\textbf
  {\bibinfo {volume} {97}},\ \bibinfo {pages} {032346} (\bibinfo {year} {2018})}\BibitemShut {NoStop}%
\bibitem [{\citenamefont {Leung}\ \emph {et~al.}(1997)\citenamefont {Leung}, \citenamefont {Nielsen}, \citenamefont {Chuang},\ and\ \citenamefont {Yamamoto}}]{Leung1997}%
  \BibitemOpen
  \bibfield  {author} {\bibinfo {author} {\bibfnamefont {D.~W.}\ \bibnamefont {Leung}}, \bibinfo {author} {\bibfnamefont {M.~A.}\ \bibnamefont {Nielsen}}, \bibinfo {author} {\bibfnamefont {I.~L.}\ \bibnamefont {Chuang}},\ and\ \bibinfo {author} {\bibfnamefont {Y.}~\bibnamefont {Yamamoto}},\ }\bibfield  {title} {\bibinfo {title} {Approximate quantum error correction can lead to better codes},\ }\href {https://doi.org/10.1103/PhysRevA.56.2567} {\bibfield  {journal} {\bibinfo  {journal} {Phys. Rev. A}\ }\textbf {\bibinfo {volume} {56}},\ \bibinfo {pages} {2567} (\bibinfo {year} {1997})}\BibitemShut {NoStop}%
\bibitem [{\citenamefont {Ng}\ and\ \citenamefont {Mandayam}(2010)}]{Ng2010}%
  \BibitemOpen
  \bibfield  {author} {\bibinfo {author} {\bibfnamefont {H.~K.}\ \bibnamefont {Ng}}\ and\ \bibinfo {author} {\bibfnamefont {P.}~\bibnamefont {Mandayam}},\ }\bibfield  {title} {\bibinfo {title} {Simple approach to approximate quantum error correction based on the transpose channel},\ }\href {https://doi.org/10.1103/PhysRevA.81.062342} {\bibfield  {journal} {\bibinfo  {journal} {Phys. Rev. A}\ }\textbf {\bibinfo {volume} {81}},\ \bibinfo {pages} {062342} (\bibinfo {year} {2010})}\BibitemShut {NoStop}%
\bibitem [{\citenamefont {B\'eny}\ and\ \citenamefont {Oreshkov}(2010)}]{Beny2010}%
  \BibitemOpen
  \bibfield  {author} {\bibinfo {author} {\bibfnamefont {C.}~\bibnamefont {B\'eny}}\ and\ \bibinfo {author} {\bibfnamefont {O.}~\bibnamefont {Oreshkov}},\ }\bibfield  {title} {\bibinfo {title} {General conditions for approximate quantum error correction and near-optimal recovery channels},\ }\href {https://doi.org/10.1103/PhysRevLett.104.120501} {\bibfield  {journal} {\bibinfo  {journal} {Phys. Rev. Lett.}\ }\textbf {\bibinfo {volume} {104}},\ \bibinfo {pages} {120501} (\bibinfo {year} {2010})}\BibitemShut {NoStop}%
\bibitem [{\citenamefont {Zheng}\ \emph {et~al.}(2024)\citenamefont {Zheng}, \citenamefont {He}, \citenamefont {Lee},\ and\ \citenamefont {Jiang}}]{Zheng2024}%
  \BibitemOpen
  \bibfield  {author} {\bibinfo {author} {\bibfnamefont {G.}~\bibnamefont {Zheng}}, \bibinfo {author} {\bibfnamefont {W.}~\bibnamefont {He}}, \bibinfo {author} {\bibfnamefont {G.}~\bibnamefont {Lee}},\ and\ \bibinfo {author} {\bibfnamefont {L.}~\bibnamefont {Jiang}},\ }\bibfield  {title} {\bibinfo {title} {Near-optimal performance of quantum error correction codes},\ }\href {https://doi.org/10.1103/physrevlett.132.250602} {\bibfield  {journal} {\bibinfo  {journal} {Physical Review Letters}\ }\textbf {\bibinfo {volume} {132}},\ \bibinfo {pages} {250602} (\bibinfo {year} {2024})}\BibitemShut {NoStop}%
\bibitem [{\citenamefont {Li}\ \emph {et~al.}(2025)\citenamefont {Li}, \citenamefont {Wang}, \citenamefont {Zheng}, \citenamefont {Wong},\ and\ \citenamefont {Jiang}}]{Li2025}%
  \BibitemOpen
  \bibfield  {author} {\bibinfo {author} {\bibfnamefont {B.}~\bibnamefont {Li}}, \bibinfo {author} {\bibfnamefont {Z.}~\bibnamefont {Wang}}, \bibinfo {author} {\bibfnamefont {G.}~\bibnamefont {Zheng}}, \bibinfo {author} {\bibfnamefont {Y.}~\bibnamefont {Wong}},\ and\ \bibinfo {author} {\bibfnamefont {L.}~\bibnamefont {Jiang}},\ }\bibfield  {title} {\bibinfo {title} {Optimality condition for the petz map},\ }\href {https://doi.org/10.1103/PhysRevLett.134.200602} {\bibfield  {journal} {\bibinfo  {journal} {Phys. Rev. Lett.}\ }\textbf {\bibinfo {volume} {134}},\ \bibinfo {pages} {200602} (\bibinfo {year} {2025})}\BibitemShut {NoStop}%
\bibitem [{\citenamefont {Nielsen}\ and\ \citenamefont {Chuang}(2010)}]{Nielsen_Chuang_2010}%
  \BibitemOpen
  \bibfield  {author} {\bibinfo {author} {\bibfnamefont {M.~A.}\ \bibnamefont {Nielsen}}\ and\ \bibinfo {author} {\bibfnamefont {I.~L.}\ \bibnamefont {Chuang}},\ }\href@noop {} {\emph {\bibinfo {title} {Quantum Computation and Quantum Information: 10th Anniversary Edition}}}\ (\bibinfo  {publisher} {Cambridge University Press},\ \bibinfo {year} {2010})\BibitemShut {NoStop}%
\bibitem [{\citenamefont {Jiang}\ \emph {et~al.}(2020)\citenamefont {Jiang}, \citenamefont {Kathuria}, \citenamefont {Lee}, \citenamefont {Padmanabhan},\ and\ \citenamefont {Song}}]{Jiang2020}%
  \BibitemOpen
  \bibfield  {author} {\bibinfo {author} {\bibfnamefont {H.}~\bibnamefont {Jiang}}, \bibinfo {author} {\bibfnamefont {T.}~\bibnamefont {Kathuria}}, \bibinfo {author} {\bibfnamefont {Y.~T.}\ \bibnamefont {Lee}}, \bibinfo {author} {\bibfnamefont {S.}~\bibnamefont {Padmanabhan}},\ and\ \bibinfo {author} {\bibfnamefont {Z.}~\bibnamefont {Song}},\ }\bibfield  {title} {\bibinfo {title} {A faster interior point method for semidefinite programming},\ }\bibfield  {journal} {\bibinfo  {journal} {IEEE}\ }\href {https://doi.org/10.1109/FOCS46700.2020.00089} {10.1109/FOCS46700.2020.00089} (\bibinfo {year} {2020}), \BibitemShut {NoStop}%
\bibitem [{\citenamefont {Schwinger}(1952)}]{Schwinger1952}%
  \BibitemOpen
  \bibfield  {author} {\bibinfo {author} {\bibfnamefont {J.}~\bibnamefont {Schwinger}},\ }\bibfield  {title} {\bibinfo {title} {On angular momentum},\ }\href {https://www.ifi.unicamp.br/~cabrera/teaching/paper_schwinger.pdf} {\bibfield  {journal} {\bibinfo  {journal} {Unpublished}\ } (\bibinfo {year} {1952})}\BibitemShut {NoStop}%
\bibitem [{\citenamefont {Dubus}\ \emph {et~al.}(2024)\citenamefont {Dubus}, \citenamefont {Haas},\ and\ \citenamefont {Cerf}}]{Dubus2024}%
  \BibitemOpen
  \bibfield  {author} {\bibinfo {author} {\bibfnamefont {B.}~\bibnamefont {Dubus}}, \bibinfo {author} {\bibfnamefont {T.}~\bibnamefont {Haas}},\ and\ \bibinfo {author} {\bibfnamefont {N.~J.}\ \bibnamefont {Cerf}},\ }\href {https://doi.org/10.48550/ARXIV.2411.04918} {\bibinfo {title} {From bosons and fermions to spins: A multi-mode extension of the jordan-schwinger map}} (\bibinfo {year} {2024}),\ \Eprint {https://arxiv.org/abs/2411.04918} {arXiv:2411.04918 [quant-ph]} \BibitemShut {NoStop}%
\bibitem [{\citenamefont {Kraus}\ \emph {et~al.}(1983)\citenamefont {Kraus}, \citenamefont {Böhm}, \citenamefont {Dollard},\ and\ \citenamefont {Wootters}}]{Kraus1983}%
  \BibitemOpen
  \bibinfo {editor} {\bibfnamefont {K.}~\bibnamefont {Kraus}}, \bibinfo {editor} {\bibfnamefont {A.}~\bibnamefont {Böhm}}, \bibinfo {editor} {\bibfnamefont {J.~D.}\ \bibnamefont {Dollard}},\ and\ \bibinfo {editor} {\bibfnamefont {W.~H.}\ \bibnamefont {Wootters}},\ eds.,\ \href {https://doi.org/10.1007/3-540-12732-1} {\emph {\bibinfo {title} {States, Effects, and Operations: Fundamental Notions of Quantum Theory}}},\ \bibinfo {edition} {1st}\ ed.,\ Lecture Notes in Physics\ (\bibinfo  {publisher} {Springer Berlin Heidelberg},\ \bibinfo {year} {1983})\BibitemShut {NoStop}%
\bibitem [{\citenamefont {Chuang}\ \emph {et~al.}(1997)\citenamefont {Chuang}, \citenamefont {Leung},\ and\ \citenamefont {Yamamoto}}]{Chuang1997}%
  \BibitemOpen
  \bibfield  {author} {\bibinfo {author} {\bibfnamefont {I.~L.}\ \bibnamefont {Chuang}}, \bibinfo {author} {\bibfnamefont {D.~W.}\ \bibnamefont {Leung}},\ and\ \bibinfo {author} {\bibfnamefont {Y.}~\bibnamefont {Yamamoto}},\ }\bibfield  {title} {\bibinfo {title} {Bosonic quantum codes for amplitude damping},\ }\href {https://doi.org/10.1103/physreva.56.1114} {\bibfield  {journal} {\bibinfo  {journal} {Physical Review A}\ }\textbf {\bibinfo {volume} {56}},\ \bibinfo {pages} {1114} (\bibinfo {year} {1997})}\BibitemShut {NoStop}%
\bibitem [{\citenamefont {Dodonov}\ \emph {et~al.}(1974)\citenamefont {Dodonov}, \citenamefont {Malkin},\ and\ \citenamefont {Man’ko}}]{Dodonov1974}%
  \BibitemOpen
  \bibfield  {author} {\bibinfo {author} {\bibfnamefont {V.}~\bibnamefont {Dodonov}}, \bibinfo {author} {\bibfnamefont {I.}~\bibnamefont {Malkin}},\ and\ \bibinfo {author} {\bibfnamefont {V.}~\bibnamefont {Man’ko}},\ }\bibfield  {title} {\bibinfo {title} {Even and odd coherent states and excitations of a singular oscillator},\ }\href {https://doi.org/10.1016/0031-8914(74)90215-8} {\bibfield  {journal} {\bibinfo  {journal} {Physica}\ }\textbf {\bibinfo {volume} {72}},\ \bibinfo {pages} {597} (\bibinfo {year} {1974})}\BibitemShut {NoStop}%
\bibitem [{\citenamefont {Slepian}(1968)}]{Slepian1968}%
  \BibitemOpen
  \bibfield  {author} {\bibinfo {author} {\bibfnamefont {D.}~\bibnamefont {Slepian}},\ }\bibfield  {title} {\bibinfo {title} {Group codes for the gaussian channel},\ }\href {https://doi.org/10.1002/j.1538-7305.1968.tb02486.x} {\bibfield  {journal} {\bibinfo  {journal} {Bell System Technical Journal}\ }\textbf {\bibinfo {volume} {47}},\ \bibinfo {pages} {575} (\bibinfo {year} {1968})}\BibitemShut {NoStop}%
\bibitem [{\citenamefont {Ericson}\ and\ \citenamefont {Zinoviev}(2001)}]{ericson2001}%
  \BibitemOpen
  \bibfield  {author} {\bibinfo {author} {\bibfnamefont {T.}~\bibnamefont {Ericson}}\ and\ \bibinfo {author} {\bibfnamefont {V.}~\bibnamefont {Zinoviev}},\ }\href {https://books.google.ca/books?id=Im0a0gEACAAJ} {\emph {\bibinfo {title} {Codes on Euclidean Spheres}}},\ North-Holland mathematical library\ (\bibinfo  {publisher} {Elsevier},\ \bibinfo {year} {2001})\BibitemShut {NoStop}%
\bibitem [{\citenamefont {Shannon}(1949)}]{Shannon1949}%
  \BibitemOpen
  \bibfield  {author} {\bibinfo {author} {\bibfnamefont {C.}~\bibnamefont {Shannon}},\ }\bibfield  {title} {\bibinfo {title} {Communication in the presence of noise},\ }\href {https://doi.org/10.1109/jrproc.1949.232969} {\bibfield  {journal} {\bibinfo  {journal} {Proceedings of the IRE}\ }\textbf {\bibinfo {volume} {37}},\ \bibinfo {pages} {10} (\bibinfo {year} {1949})}\BibitemShut {NoStop}%
\bibitem [{\citenamefont {Mirrahimi}\ \emph {et~al.}(2014)\citenamefont {Mirrahimi}, \citenamefont {Leghtas}, \citenamefont {Albert}, \citenamefont {Touzard}, \citenamefont {Schoelkopf}, \citenamefont {Jiang},\ and\ \citenamefont {Devoret}}]{Mirrahimi2014}%
  \BibitemOpen
  \bibfield  {author} {\bibinfo {author} {\bibfnamefont {M.}~\bibnamefont {Mirrahimi}}, \bibinfo {author} {\bibfnamefont {Z.}~\bibnamefont {Leghtas}}, \bibinfo {author} {\bibfnamefont {V.~V.}\ \bibnamefont {Albert}}, \bibinfo {author} {\bibfnamefont {S.}~\bibnamefont {Touzard}}, \bibinfo {author} {\bibfnamefont {R.~J.}\ \bibnamefont {Schoelkopf}}, \bibinfo {author} {\bibfnamefont {L.}~\bibnamefont {Jiang}},\ and\ \bibinfo {author} {\bibfnamefont {M.~H.}\ \bibnamefont {Devoret}},\ }\bibfield  {title} {\bibinfo {title} {Dynamically protected cat-qubits: a new paradigm for universal quantum computation},\ }\href {https://doi.org/10.1088/1367-2630/16/4/045014} {\bibfield  {journal} {\bibinfo  {journal} {New Journal of Physics}\ }\textbf {\bibinfo {volume} {16}},\ \bibinfo {pages} {045014} (\bibinfo {year} {2014})}\BibitemShut {NoStop}%
\bibitem [{\citenamefont {Wetherbee}\ \emph {et~al.}(2026)\citenamefont {Wetherbee}, \citenamefont {Xu}, \citenamefont {Albert}, \citenamefont {Royer},\ and\ \citenamefont {Fatemi}}]{Wetherbee2026}%
  \BibitemOpen
  \bibfield  {author} {\bibinfo {author} {\bibfnamefont {O.~C.}\ \bibnamefont {Wetherbee}}, \bibinfo {author} {\bibfnamefont {Y.}~\bibnamefont {Xu}}, \bibinfo {author} {\bibfnamefont {V.~V.}\ \bibnamefont {Albert}}, \bibinfo {author} {\bibfnamefont {B.}~\bibnamefont {Royer}},\ and\ \bibinfo {author} {\bibfnamefont {V.}~\bibnamefont {Fatemi}},\ }\href {https://doi.org/10.48550/ARXIV.2606.11010} {\bibinfo {title} {Bosonic cyclic codes: Trading stabilizers for gaussian non-clifford phase gates}} (\bibinfo {year} {2026}),\ \Eprint {https://arxiv.org/abs/2606.11010} {arXiv:2606.11010 [quant-ph]} \BibitemShut {NoStop}%
\bibitem [{\citenamefont {Albert}\ \emph {et~al.}(2019)\citenamefont {Albert}, \citenamefont {Mundhada}, \citenamefont {Grimm}, \citenamefont {Touzard}, \citenamefont {Devoret},\ and\ \citenamefont {Jiang}}]{Albert2019}%
  \BibitemOpen
  \bibfield  {author} {\bibinfo {author} {\bibfnamefont {V.~V.}\ \bibnamefont {Albert}}, \bibinfo {author} {\bibfnamefont {S.~O.}\ \bibnamefont {Mundhada}}, \bibinfo {author} {\bibfnamefont {A.}~\bibnamefont {Grimm}}, \bibinfo {author} {\bibfnamefont {S.}~\bibnamefont {Touzard}}, \bibinfo {author} {\bibfnamefont {M.~H.}\ \bibnamefont {Devoret}},\ and\ \bibinfo {author} {\bibfnamefont {L.}~\bibnamefont {Jiang}},\ }\bibfield  {title} {\bibinfo {title} {Pair-cat codes: autonomous error-correction with low-order nonlinearity},\ }\href {https://doi.org/10.1088/2058-9565/ab1e69} {\bibfield  {journal} {\bibinfo  {journal} {Quantum Science and Technology}\ }\textbf {\bibinfo {volume} {4}},\ \bibinfo {pages} {035007} (\bibinfo {year} {2019})}\BibitemShut {NoStop}%
\bibitem [{\citenamefont {Biswas}\ \emph {et~al.}(2026)\citenamefont {Biswas}, \citenamefont {Sharma}, \citenamefont {Salvador}, \citenamefont {Wang}, \citenamefont {Granath}, \citenamefont {Udupa},\ and\ \citenamefont {Ferrini}}]{Biswas2026}%
  \BibitemOpen
  \bibfield  {author} {\bibinfo {author} {\bibfnamefont {D.}~\bibnamefont {Biswas}}, \bibinfo {author} {\bibfnamefont {N.}~\bibnamefont {Sharma}}, \bibinfo {author} {\bibfnamefont {A.}~\bibnamefont {Salvador}}, \bibinfo {author} {\bibfnamefont {R.}~\bibnamefont {Wang}}, \bibinfo {author} {\bibfnamefont {M.}~\bibnamefont {Granath}}, \bibinfo {author} {\bibfnamefont {A.}~\bibnamefont {Udupa}},\ and\ \bibinfo {author} {\bibfnamefont {G.}~\bibnamefont {Ferrini}},\ }\href {https://arxiv.org/abs/2607.00786} {\bibinfo {title} {Bias-preserving gates and quantum error correction with dual-rail cat codes}} (\bibinfo {year} {2026}),\ \Eprint {https://arxiv.org/abs/2607.00786} {arXiv:2607.00786 [quant-ph]} \BibitemShut {NoStop}%
\bibitem [{\citenamefont {Guillaud}\ and\ \citenamefont {Mirrahimi}(2019)}]{Guillaud2019}%
  \BibitemOpen
  \bibfield  {author} {\bibinfo {author} {\bibfnamefont {J.}~\bibnamefont {Guillaud}}\ and\ \bibinfo {author} {\bibfnamefont {M.}~\bibnamefont {Mirrahimi}},\ }\bibfield  {title} {\bibinfo {title} {Repetition cat qubits for fault-tolerant quantum computation},\ }\href {https://doi.org/10.1103/physrevx.9.041053} {\bibfield  {journal} {\bibinfo  {journal} {Physical Review X}\ }\textbf {\bibinfo {volume} {9}},\ \bibinfo {pages} {041053} (\bibinfo {year} {2019})}\BibitemShut {NoStop}%
\bibitem [{\citenamefont {Kubischta}\ and\ \citenamefont {Teixeira}(2024)}]{Kubischta2024}%
  \BibitemOpen
  \bibfield  {author} {\bibinfo {author} {\bibfnamefont {E.}~\bibnamefont {Kubischta}}\ and\ \bibinfo {author} {\bibfnamefont {I.}~\bibnamefont {Teixeira}},\ }\bibfield  {title} {\bibinfo {title} {Quantum codes from twisted unitary $t$-groups},\ }\href {https://doi.org/10.1103/physrevlett.133.030602} {\bibfield  {journal} {\bibinfo  {journal} {Phys. Rev. Lett. 133, 030602 (2024)}\ }\textbf {\bibinfo {volume} {133}},\ \bibinfo {pages} {030602} (\bibinfo {year} {2024})},\ \Eprint {https://arxiv.org/abs/2402.01638} {arXiv:2402.01638 [quant-ph]} \BibitemShut {NoStop}%
\end{thebibliography}
%

\end{document}